\documentclass[aps,pre,reprint,floatfix,showkeys,longbibliography]{revtex4-1}	
\usepackage[utf8]{inputenc}     
\usepackage[T1]{fontenc}
\usepackage[english]{babel}   

\usepackage{lipsum}
\usepackage[hidelinks,colorlinks=true]{hyperref}

\usepackage[usenames,dvipsnames,svgnames,table]{xcolor}
\usepackage{graphicx}
\usepackage{mathbbol}

\definecolor{myred}{RGB}{228,26,28}
\definecolor{myblue}{RGB}{55,126,184}
\definecolor{myorange}{RGB}{225,127,0}
\definecolor{mygreen}{RGB}{77,175,74}
\definecolor{mylila}{RGB}{152,78,163}
\definecolor{mybrown}{RGB}{153,76,0}
\definecolor{mygray}{RGB}{153,153,153}
\definecolor{darkred}{rgb}{0.8,0,0}
\definecolor{mydarkgreen}{RGB}{0,102,0}
\definecolor{mydarkbrown}{RGB}{102,52,0}
\definecolor{Orange}{RGB}{235,129,27}
\definecolor{Green}{RGB}{35,55,59}
\usepackage{bm}
\usepackage{amsmath,amsfonts,amssymb}

\usepackage{tikz} 
\usetikzlibrary{decorations,backgrounds,patterns} 
\usepackage{pgfplots} 
\usepackage{pgfplotstable}

\usepgfplotslibrary{external}
\pgfplotsset{compat=newest}
\usetikzlibrary{pgfplots.groupplots}
\usetikzlibrary{calc}
\usetikzlibrary{3d}
\usetikzlibrary{positioning}
\usepgfplotslibrary{units}
\pgfplotsset{unit code/.code={\si{#1}}}
\usepgfplotslibrary{patchplots}
\usepgfplotslibrary{fillbetween}
\tikzset{>=stealth}

\usetikzlibrary{decorations.markings}
\makeatletter
\tikzset{
  nomorepostactions/.code={\let\tikz@postactions=\pgfutil@empty},
  mymark/.style 2 args={decoration={markings,
    mark= between positions 0 and 1 step (1/30)*\pgfdecoratedpathlength with{%
        \tikzset{#2,every mark}\tikz@options
        \pgfuseplotmark{#1}%
      },  
    },
    postaction={decorate},
    /pgfplots/legend image post style={
        mark=#1,mark options={#2},every path/.append style={nomorepostactions}
    },
  },
}
\makeatother

\newcommand{\idest}{\textsl{i.e. }}

\renewcommand{\d}{\mathrm{d}}
\newcommand{\D}{\partial}

\newcommand{\euler}{\mathrm{e}}
\renewcommand{\i}{\mathrm{i}}

\newcommand{\covariance}{\Upsilon}

\newcommand{\pos}{x}
\newcommand{\velocity}{v}
\newcommand{\coordinate}{\Gamma}
\newcommand{\potential}{V}
\newcommand{\force}{g}
\newcommand{\temperature}{T}
\newcommand{\invtemperature}{\beta}
\newcommand{\mass}{m}
\newcommand{\noise}{\eta}
\newcommand{\friction}{\xi}
\newcommand{\probability}{P}

\newcommand{\currentvec}{\bm{J}}
\newcommand{\energydensity}{e}
\newcommand{\energy}{E}
\newcommand{\heat}{Q}
\newcommand{\work}{W}
\newcommand{\entropydensity}{s}
\newcommand{\entropy}{S}
\newcommand{\ep}{\Sigma}
\newcommand{\freeenergydensity}{f}
\newcommand{\freeenergy}{F}
\newcommand{\mutual}{I}
\newcommand{\hamiltonian}{H}

\begin{document}

\title{Effective thermodynamics of two interacting underdamped Brownian particles}

\author{Tim Herpich}
\email{tim.herpich@uni.lu}	
\author{Kamran Shayanfard}
\author{Massimiliano Esposito}
\email{massimiliano.esposito@uni.lu}	
\affiliation{Complex Systems and Statistical Mechanics, Physics and Materials Science Research Unit, University of Luxembourg, L-1511 Luxembourg, Luxembourg}
\date{\today}

\begin{abstract}
Starting from the stochastic thermodynamics description of two coupled underdamped Brownian particles, we showcase and compare three different coarse-graining schemes leading to an effective thermodynamic description for the first of the two particles: marginalization over one particle, bipartite structure with information flows and the Hamiltonian of mean force formalism.
In the limit of time-scale separation where the second particle with a fast relaxation time scale locally equilibrates with respect to the coordinates of the first slowly relaxing particle, the effective thermodynamics resulting from the first and third approach are shown to capture the full thermodynamics and to coincide with each other.
In the bipartite approach, the slow part does not, in general, allow for an exact thermodynamic description as the entropic exchange between the particles is ignored.
Physically, the second particle effectively becomes part of the heat reservoir.
In the limit where the second particle becomes heavy and thus deterministic, the effective thermodynamics of the first two coarse-graining methods coincides with the full one.
The Hamiltonian of mean force formalism however is shown to be incompatible with that limit.
Physically, the second particle becomes a work source.
These theoretical results are illustrated using an exactly solvable harmonic model. 
\end{abstract}

\maketitle

\section{Introduction}   \label{sec:introduction}

Over the past two decades, stochastic thermodynamics established the tools to formulate thermodynamics for small systems subjected to significant fluctuations and driven far from equilibrium \cite{seifert2012rpp,broeck2015physica,sekimoto2010,zhang2012pr,ge2012pr,jarzynski2010ar,rao2018njp}. This theory has been successful in various contexts, \emph{e.g.}, Brownian particles \cite{ciliberto2017prx, proesmans2016prx}, electronic systems \cite{pekkola2013rmp}, chemical reaction networks \cite{rao2016prx,rao2018jcp}, active matter \cite{cates2017prx,eichhorn2019prx} and information processing \cite{parrondo2015np}. In a nutshell, stochastic thermodynamics consistently builds a thermodynamic structure on top of a stochastic process described by master equations \cite{vandenbroeck2010pre} or Fokker-Planck equations \cite{vandenbroeck2010pre2}, implicitly assuming that the traced out degrees of freedom always stay at equilibrium.

In this paper we want to address two apparently distinct questions within the framework of underdamped Fokker-Planck dynamics.
First, we want to shed light on the nature of heat and work by understanding how a subset of degrees of freedom from the system can start to behave as a thermal bath or a work source, respectively.
For systems characterized by master equations, it was proven that if there is a time-scale separation (TSS) between the slow and the fast degrees of freedom, the latter equilibrate with respect to the slow coordinates and represent an ideal heat reservoir the slow degrees of freedom are coupled with \cite{esposito2012pre}.
Instead, the conditions under which a subset of degrees of freedom can generate a stochastic driving on the system energies that can be treated as a work source have been identified in Ref. \cite{verley2014njp}. However, the limit of a smooth deterministic driving of the energies requires a limit that only an underdamped Fokker-Planck equation can provide. We aim therefore at reconsidering these questions in this paper within the framework of underdamped Brownian dynamics.
Secondly, we want to consider various coarse-graining schemes preserving thermodynamic consistency that have been proposed in the literature
\cite{altaner2012prl,aurell2012prl,cuetara2011prb,wachtel2018njp,seifert2012prl,herpich2018prx,herpich2019pre,shengentropy2019,sekimoto2007pre,parrondo2008pre,seifert2013jcp,polettini2017jsm,speck2015njp,parrondo2018nc,kahlen2018jsm,seifertjsm2018,gupta2018jsm}.
In particular, we want to focus on three different well-established approaches that have been considered for stochastic dynamics governed by master equations.
First, we consider the most straightforward approach where a subset of states is explicitly coarse-grained and the effective thermodynamics is defined for that reduced dynamics as one formally would for the full dynamics \cite{esposito2012pre,bo2014jsp}.
Next, we consider an approach based on splitting the full system in two parts resulting in effective second laws for each parts which are modified by a term describing the transfer of mutual information between each parts. This approach provides a convenient framework to describe how a Maxwell demon \cite{leff1990} mechanism can produce an information flow that is consumed by the system to drive processes against their spontaneous direction \cite{esposito2014prx,horowitz2014njp,horowitz2015jstatmech,hartrich2014jsm}.
Finally, we consider the so-called Hamiltonian of mean force (HMF) approach which introduces a notion of energy for a system strongly coupled to its environment \cite{seifert2016prl,jarzynski2017prx,strasberg2017pre}. 
In this paper, we will consider these various coarse-grainings for underdamped Brownian particles and discuss how they are related.
As we will see, far from being distinct, the question of the connection between the different coarse-graining schemes will provide us with a good framework to get insight into the nature of heat and work.

To achieve these goals, we will consider two coupled underdamped particles as this model already contains the key ingredients to generalize to multiple underdamped particles.
Besides interesting formal connections between entropic contributions appearing in the different coarse-graining schemes, we will find that the effective thermodynamics based on marginalization and the Hamiltonian of mean force become equivalent and capture the correct global thermodynamics in the limit of time-scale separation. In this limit, the second particle is so much faster than the first one that it instantaneously relaxes to a local equilibrium corresponding to the coordinates of the first particle.
Conversely, the thermodynamics based on the slow part of the bipartite structure does not agree with the full thermodynamics. The mismatch corresponds to the entropic contribution due to the coupling of the second particle. Physically, the coarse-grained particle becomes part of the heat reservoir. Moreover, in the limit where one particle has an exceedingly large mass compared to the other one, we will find that the former becomes a work source acting on the latter. In that case, the effective thermodynamics emerging from the first two coarse-graining schemes, marginalization and bipartite structure, again captures the correct global thermodynamics (at least up to a trivial macroscopic friction term in the work source).
In contrast, we will show that the Hamiltonian is incompatible with that limit.
These theoretical predictions will be confirmed using an analytically tractable model made up of two linearly coupled harmonic oscillators.

The plan of the paper is as follows. In Sec. \ref{sec:theory} the stochastic thermodynamics for both a single underdamped particle and two interacting and underdamped particles is formulated.
Next, in Sec. \ref{sec:coarsegraining} we formulate and compare the three different coarse-graining approaches - marginalization, bipartite perspective and Hamiltonian of mean force - for our underdamped two-particle system. The respective effective thermodynamic description is furthermore compared with the full one in the two aforementioned limits. As an example, we consider an analytically solvable model in Sec. \ref{sec:example}.
We conclude with an outlook to potential future works in Sec. \ref{sec:conclusion}.
%We conclude with a generalization of the underdamped two-body model to arbitrarily many underdamped particles in the appendix \ref{sec:appendix}.

\section{Stochastic Thermodynamics}  \label{sec:theory}
\subsection{Single Underdamped Particle} \label{sec:singletheory}

We consider a particle of mass $\mass$ with the phase-space coordinate $\bm{\coordinate}=(\bm{\pos},\bm{\velocity})^\top \in  \mathbb{R}^6$, where $\bm{\pos} \in  \mathbb{R}^3$ and $\bm{\velocity} \in  \mathbb{R}^3$ denote position and velocity of the particle, respectively. The particle moves in a time-dependent potential $\potential(\bm{\pos},t)$, hence its Hamiltonian reads
\begin{align} \label{eq:singleenergydensity}
\energydensity(\bm{\coordinate},t) =  \frac{\mass}{2} \bm{\velocity}^2 + \potential(\bm{\pos},t) .
\end{align}
The particle is furthermore subjected to a generic force $\bm{\force}(\bm{\coordinate},t)$. If the force is conservative it derives from a potential, $\bm{\force}(\bm{\pos},t) ~=~ - \D_{\bm{\pos}} \hat{\potential}(\bm{\pos},t)$. In this case, one can define heat and work in different ways depending on whether $ \hat{\potential}(\bm{\pos},t)$ is part of the system Hamiltonian \eqref{eq:singleenergydensity} or not, see Ref. \cite{Jarzynski07b}. Conversely, if the force $\bm{\force}$ is nonconservative, it does not derive from a potential. For generality, and since it will be useful later, we assume that the force may be velocity-dependent, $\bm{\force}(\bm{\coordinate},t)$.

The system is coupled to a heat reservoir at inverse temperature $\invtemperature$, giving rise to zero-mean delta-correlated Gaussian white noise
\begin{align}\label{eq:noise}
\langle \bm{\noise}_i(t) \rangle = 0, \quad \langle \bm{\noise}_i(t) \bm{\noise}_j(t') \rangle = 2 \friction \, \invtemperature^{-1} \, \delta_{ij} \delta(t-t') ,
\end{align}
for $i,j=1,2,3$. We denote by $\xi$ the friction the particle experiences and set $k_B \equiv 1$ in the following.
Then, the stochastic dynamics of the system is governed by the following Langevin equation
\begin{align}\label{eq:singlelangevin}
\begin{pmatrix}
\dot{\bm{\pos}} \\ \dot{\bm{\velocity}}
\end{pmatrix} =
\begin{pmatrix}
\bm{\velocity} \\ \tfrac{1}{\mass} \left[ -\D_{\bm{\pos}} \potential(\bm{\pos},t) + \bm{\force}(\bm{\coordinate},t) - \friction \, \bm{\velocity} + \bm{\noise}(t) \right]
\end{pmatrix},
\end{align}
and the equivalent Fokker-Planck equation ruling the time evolution of the probability density $\probability(\bm{\coordinate},t)$ reads
\begin{align} \label{eq:singlefokkerplanckone}
\D_t \, \probability = - \nabla \cdot ( \bm{\mu} \probability ) + \nabla \cdot \big( \bm{D} \cdot \nabla \probability \big)  ,
\end{align}
with the drift and diffusion matrices
\begin{align}
\bm{\mu} &= \begin{pmatrix}
\bm{\velocity} \\ \tfrac{1}{\mass} \left[ -\D_{\bm{\pos}} \potential(\bm{\pos},t) + \bm{\force}(\bm{\coordinate},t) - \friction \, \bm{\velocity}  \right]
\end{pmatrix} \\ 
\bm{D}_{ij} &= \frac{\friction \, \delta_{i j} }{\invtemperature \mass^2} \sum\limits_{n=4}^6  \delta_{i n} \, ,
\end{align}
and the nabla operator $\nabla \equiv (\D_{\bm{\pos}},\D_{\bm{\velocity}})^{\top} $. The Fokker-Planck Eq. \eqref{eq:singlefokkerplanckone} can be cast into a continuity equation
\begin{align}\label{eq:singlefokkerplanck}
\D_t \probability = - \nabla \cdot \currentvec = -\nabla \cdot \left( \bm{L}^{det} + \bm{L}^{diss} \right) \probability .
\end{align}
Here, the probability current $\currentvec$ is split into a deterministic contribution 
\begin{align}\label{eq:singlefokkerplanckdeterministiccurrent}
\bm{L}^{det} = 
\begin{pmatrix} 
\bm{\velocity} \\ \tfrac{1}{\mass} \left[ - \D_{\bm{\pos}} \potential(\bm{\pos},t) + \bm{\force}(\bm{\coordinate},t) \right] 
\end{pmatrix}  ,
\end{align} 
and a dissipative one 
\begin{align}\label{eq:singlefokkerplanckdissipativecurrent}
\bm{L}^{diss} =  - \frac{\friction}{\mass^2} 
\begin{pmatrix} 
0 \\ \mass \bm{\velocity} + \invtemperature^{-1} \, \D_{\bm{\velocity}} \ln \probability 
\end{pmatrix}.
\end{align} 
The average energy of the particle is 
\begin{align}\label{eq:singleaverageeneergy} 
\energy = \int \d \bm{\coordinate} \; \energydensity \, \probability ,
\end{align} 
and its rate of change 
\begin{align}\label{eq:singlefirstlaw}
d_t \energy = \dot{\heat} + \dot{\work}, 
\end{align} 
can be decomposed into a work current
\begin{align}\label{eq:singlework}
\dot{\work} = \int \d \bm{\coordinate} \; \probability \, \D_t \energydensity + \int \d \bm{\coordinate} \; \probability \, \bm{\force} \cdot \bm{\velocity} ,
\end{align}
and into a heat current
\begin{align}\label{eq:singleheat}
\dot{\heat} = \int \d \bm{\coordinate} \; \energydensity \, \D_t \probability - \int \d \bm{\coordinate} \; \probability \, \bm{\force} \cdot \bm{\velocity} .
\end{align}
Eq. \eqref{eq:singlefirstlaw} constitutes the first law of thermodynamics ensuring energy conservation \cite{sekimoto1997ptps}.
Using Eq. \eqref{eq:singlefokkerplanck}, the heat current can be written as follows
\begin{align} \label{eq:singleheatcurrent}
\dot{\heat} = - \friction \int \d \bm{\coordinate} \; \probability \, \left(\bm{\velocity} + \frac{1}{\invtemperature \mass} \D_{\bm{\velocity}} \ln \probability \right)  \bm{\velocity} \, .
\end{align}
The nonequilibrium system entropy associated with the particle at $\bm{\coordinate}$ is defined as \cite{seifert2005prl}
\begin{align} \label{eq:singleentropydensity}
\entropydensity(\bm{\coordinate}) = -\ln \probability ,
\end{align}
where the ensemble average coincides with the Shannon entropy 
\begin{align} \label{eq:singleentropy}
\entropy = - \int \d \bm{\coordinate} \; \probability \, \ln \probability . 
\end{align}
Its time-derivative
\begin{align} 
\d_t \entropy &= \int \d \bm{\coordinate} \, [ \nabla \cdot \bm{L}^{diss} ] \, \probability + \dot{\mutual}_F  =  \invtemperature \dot{\heat} + \dot{\ep} +  \dot{\mutual}_F , \label{eq:singleentropybalance}
\end{align} 
can be split into the entropy flow from the bath to the system, $\invtemperature \dot{\heat}$, and the entropy production rate
\begin{align}  \label{eq:singlesecondlaw}
\dot{\ep} = \invtemperature \, \friction \int \d \bm{\coordinate} \; \probability \, \left(\bm{\velocity} + \frac{1}{\invtemperature \mass} \D_{\bm{\velocity}} \ln \probability \right)^2 \geq 0 ,
\end{align}
whose nonnegativity constitutes the second law of thermodynamics.
Since it will be useful later, we introduced the notation
\begin{align} \label{eq:singleinformationforce}
\dot{\mutual}_F \equiv \frac{1}{\mass} \int \d \bm{\coordinate} \; \probability \, \D_{\bm{\velocity}} \cdot \bm{\force} .
\end{align} 
Defining the nonequilibrium free-energy density $\freeenergydensity(\bm{\coordinate}) ~=~ \energydensity(\bm{\coordinate}) - \invtemperature^{-1} \entropydensity(\bm{\coordinate})$, one has for the average nonequilibrium free energy
\begin{align} \label{eq:singlefreenergy}
\freeenergy = \int \d \bm{\coordinate} \, \probability \; \freeenergydensity = \energy - \invtemperature^{-1} \entropy .
\end{align}
Eq. \eqref{eq:singlefreenergy} allows us to rewrite the work and heat current in Eqs. \eqref{eq:singlework} and \eqref{eq:singleheatcurrent} as
\begin{align} \label{eq:singleheatworkcurrent}
\begin{aligned}
\dot{\work} &= \int \d \bm{\coordinate} \; \probability \, \D_t \freeenergydensity + \int \d \bm{\coordinate} \probability \, \bm{\force} \cdot \bm{\velocity} \\
\dot{\heat} &= d_t (\freeenergy + \invtemperature^{-1} \entropy) - \dot{\work}  ,
\end{aligned}
\end{align}
and the entropy production rate in Eq. \eqref{eq:singlesecondlaw} as 
\begin{align}  \label{eq:singleentropyproduction} 
\dot{\ep} = \invtemperature ( \dot{\work} - \d_t \freeenergy ) - \dot{\mutual}_F \,  \geq 0 .
\end{align}
The additional term $ \dot{\mutual}_F $ in Eqs. \eqref{eq:singleentropybalance} and \eqref{eq:singleentropyproduction} illustrates that the presence of the velocity-dependent nonconservative force $\bm{\force}$ modifies the thermodynamics as noted in Refs. \cite{quian2004prl,quian2007pre}.

\subsection{Special Cases}
\label{sec:singlespecialcases}

\subsubsection{Standard Stochastic Thermodynamics}

Owing to the velocity-dependence of $\force(\bm{\coordinate},t$), Eqs. \eqref{eq:singleentropybalance} and \eqref{eq:singleentropyproduction} constitute a generalized entropy balance and a generalized second law, respectively. The standard thermodynamic formulation
\begin{align}\label{eq:singlestandardentropybalancesecondlaw}
\d_t \entropy = \invtemperature \dot{\heat} + \dot{\ep}, \quad \quad \temperature \dot{\ep}
= \dot{\work} - \d_t \freeenergy  \geq 0  , 
\end{align}
is recovered for velocity-independent or nonconservative Lorentz forces, that is forces that are orthogonal to the velocity, $\D_{\bm{\bm{\velocity}}} \cdot \bm{\force} = 0 $. In one dimension, this is only true for velocity-independent forces $ \D_{\velocity} \, \force = 0$.

\subsubsection{Deterministic Limit}

The dynamics is deterministic if $\friction=0$, which physically corresponds to a decoupling of the particle from the thermal reservoir. According to Eq. \eqref{eq:singleheatcurrent}, one has $\dot{\heat}=0$ and $\d_t \energy = \d_t \work $.
It follows furthermore from Eq. \eqref{eq:singlesecondlaw} that $\dot{\ep}=0$, hence it holds, using Eq. \eqref{eq:singleentropybalance}, that 
\begin{align}  \label{eq:singleentropybalancedeterministic}
\d_t \entropy = \frac{1}{\mass} \int \d \bm{\coordinate} \, \probability \; \D_{\bm{\velocity}} \cdot \bm{\force}.
\end{align}
Again, if $\bm{\force}$ is velocity-independent or a Lorentz force, the deterministic dynamics becomes Hamiltonian and the rate of entropy change is identically zero, $\d_t \entropy = 0$. In this case the second law is a triviality.

\subsubsection{Heavy Particle}

Finally, we consider the limit where the mass of the particle diverges, $\mass \to \infty$. We suppose that the conservative force scales with the mass, \idest $ \mathit{O}(\D_{\pos_i} \potential / \mass) = 1 \, \forall i$, to avoid the trivial case of a particle in a flat potential.
If $\friction$ and $\bm{\force}$ are finite, so that $\friction/\mass \to 0$ and $\bm{\force}/\mass \to 0$, one finds using Eqs. \eqref{eq:singleheatcurrent}, \eqref{eq:singleentropybalance} and \eqref{eq:singlesecondlaw} that
\begin{align}  \label{eq:singlesecondlawheavyparticle} 
\dot{\ep} = - \invtemperature \dot{\heat} = \invtemperature \friction \bm{\velocity}^2_t \geq 0, \quad \d_t \entropy = 0,
\end{align}
where $\bm{\velocity}_t$ is the solution of the deterministic Eqs. 
\begin{align}  \label{eq:singleequationofmotionheavyparticle} 
\d_t \bm{\pos}_t = \bm{\velocity}_t,  \quad \d_t \bm{\velocity}_t = - \frac{1}{\mass}  \left. \D_{\bm{\pos}} \potential(\bm{\pos},t) \right|_{\bm{\pos} = \bm{\pos}_t} .
\end{align}
According to Eq. \eqref{eq:singlesecondlawheavyparticle}, the heavy particle corresponds to the limit of macroscopic friction.

\subsection{Two Coupled Underdamped Particles} \label{sec:twotheory}

\begin{figure}[h!]
\begin{center}

\includegraphics[scale=1]{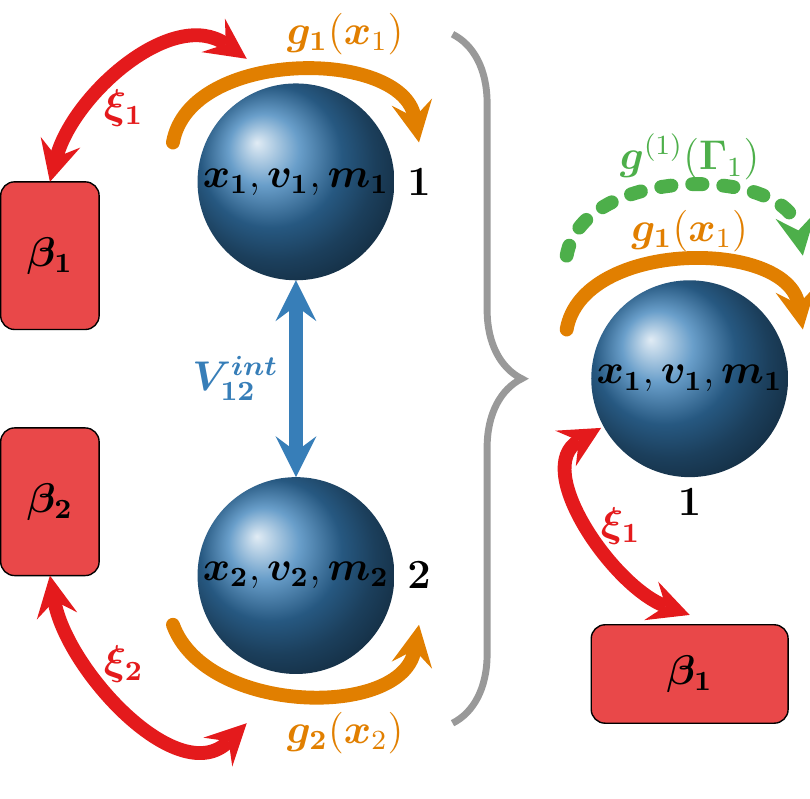}

\caption{On the left, schematics of the two underdamped and via $\potential^{int}_{12}$ interacting particles $\bm{1}$ and $\bm{2}$ that are in contact with heat reservoirs at inverse temperatures $\invtemperature_1$ and $\invtemperature_2$, respectively, are illustrated. It is furthermore assumed that both particles are subjected to nonconservative forces $\bm{\force}_i(\bm{\pos}_i)$. The right depicts the coarse-grained description of solely the first particle in the presence of an additional nonconservative force $\bm{\force}^{(1)}(\bm{\coordinate}_1)$ (green, dashed vector) that encodes the interaction with the second particle. \label{fig:modelschematics} }
\end{center}
\end{figure}

We now consider two particles labeled by $i=1,2$ of mass $\mass_i$ with the phase-space coordinate $\bm{\coordinate}_i = (\bm{\pos}_i,\bm{\velocity}_i)^\top$, as depicted in Fig. \ref{fig:modelschematics}.
The particles move in a time-dependent potential
\begin{align}   \label{eq:twopotential}
\potential(\bm{\pos}_1,\bm{\pos}_2,t) = \potential_1(\bm{\pos}_1,t) + \potential_2(\bm{\pos}_2,t) + \potential^{int}_{12} (\bm{\pos}_1,\bm{\pos}_2,t),
\end{align} 
that contains the interaction potential $\potential^{int}_{12} (\bm{\pos}_1,\bm{\pos}_2,t)$ and the Hamiltonian therefore reads
\begin{align} \label{eq:twoenergydensity}
\begin{aligned}
\energydensity(\bm{\coordinate},t) &= \frac{\mass_1}{2} \bm{\velocity}_1^2 + \frac{\mass_2}{2} \bm{\velocity}_2^2 + \potential(\bm{\pos}_1,\bm{\pos}_2,t) \\
&= \sum_i \energydensity_i(\bm{\coordinate}_i,t) + \potential^{int}_{12} (\bm{\pos}_1,\bm{\pos}_2,t)  \, ,
\end{aligned} 
\end{align} 
where we denote the bare Hamiltonian of each particle by $ \energydensity_i(\bm{\coordinate}_i,t) = \mass_i \bm{\velocity}_i^2 /2 + \potential_i(\bm{\pos}_i,t)$ with $i$=1,2. Moreover, we assume that both particles are subjected to velocity-independent nonconservative forces $\bm{\force}_i(\bm{\pos}_i,t)$ \footnote{For velocity-dependent nonconservative forces, \emph{e.g.} magnetic forces, the following procedure is analogous to the case of velocity-independent forces. The only formal modification are the additional terms, $\sum_i \! 1/\mass_i \! \int \! \d \bm{\coordinate}_i \probability_i \D_{\bm{\velocity}_i} \cdot \bm{\force}_i$, that appear in the entropy balance equation \eqref{eq:twoentropybalance}, cf. Eq. \eqref{eq:singleentropybalance}.}.

Each of the particles is connected to a heat reservoir at inverse temperature $\invtemperature_i$ giving rise to uncorrelated zero-mean Gaussian white noise 
\begin{align}   \label{eq:twolangevinnoise}
\langle \bm{\noise}^{(i)}_j(t) \rangle = 0, \quad \langle \bm{\noise}_{j}^{(i)}(t) \bm{\noise}_{j'}^{(i)}(t') \rangle = 2 \, \friction_i \invtemperature_i^{-1} \delta_{j,j'} \, \delta(t-t') ,
\end{align} 
where $\friction_i$ refers to the friction the particle $i$ experiences. The stochastic dynamics of the two-body system is ruled by the following Langevin equation
\begin{align} \label{eq:twolangevinequation}
\begin{pmatrix}
\dot{\bm{\pos}}_i \\ 
\dot{\bm{\velocity}}_i
\end{pmatrix} \!\!=\! 
\begin{pmatrix}
\bm{\velocity}_i \\ 
\!\frac{1}{\mass_i} \![ -\D_{\bm{\pos}_i} \potential(\bm{\pos}_1,\bm{\pos}_2,t) \!+\! \bm{\force}_i(\bm{\pos}_i,t) \!-\! \friction_i \bm{\velocity}_i \!+\! \bm{\noise}^{(i)}(t) ] \! 
\end{pmatrix} \! ,
\end{align}
and the equivalent Fokker-Planck equation governing the time evolution of the probability density $ \probability(\bm{\coordinate},t) $ reads 
\begin{align}  \label{eq:twofokkerplanckequation} 
\D_t \probability = - \nabla \cdot \currentvec = - \nabla \cdot \left( \bm{L}^{det} + \bm{L}^{diss} \right) \probability ,
\end{align}
with $ \nabla = ( \D_{\bm{\pos}_1} , \D_{\bm{\velocity}_1} , \D_{\bm{\pos}_2} , \D_{\bm{\velocity}_2} )^\top$. The probability current $\currentvec$ can be split into a deterministic part 
\begin{align}  \label{eq:twoprobabilitycurrentconservative}
\bm{L}^{det} = \begin{pmatrix} \bm{\velocity}_1 \\ \frac{1}{\mass_1} \left[ - \D_{\bm{\pos}_1} \potential(\bm{\pos}_1,\bm{\pos}_2,t) + \bm{\force}_1(\bm{\pos}_1,t) \right] \\ 
\bm{\velocity}_2 
\\ \frac{1}{\mass_2} \left[ - \D_{\bm{\pos}_2} \potential(\bm{\pos}_1,\bm{\pos}_2,t) + \bm{\force}_2(\bm{\pos}_2,t) \right]  \end{pmatrix}   ,
\end{align} 
and a dissipative one
\begin{align}  \label{eq:twoprobabilitycurrentdissipative}
\bm{L}^{diss}= 
\begin{pmatrix} 
0 \\ \frac{-\friction_1}{\mass_1^2} ( \mass_1 \bm{\velocity}_1 + \invtemperature_1^{-1} \D_{\bm{\velocity}_1} \ln \probability ) \\ 
0 \\ \frac{-\friction_2}{\mass_2^2} ( \mass_2 \bm{\velocity}_2 + \invtemperature_2^{-1} \D_{\bm{\velocity}_2} \ln \probability )
\end{pmatrix}.
\end{align} 
The average energy of the system is
\begin{align} \label{eq:twoaverageenergy}
\energy = \! \int \d \bm{\coordinate} \, \energydensity \, \probability ,
\end{align}
and the first law of thermodynamics reads
\begin{align} \label{eq:twofirstlaw}
\d_t \energy = \dot{\heat} + \dot{\work} , 
\end{align} 
with the heat and work current
\begin{align} \label{eq:twoheatdefintion}
\dot{\heat} &= \int \d \bm{\coordinate} \, \energydensity \, \dot{\probability}  - \int \d \bm{\coordinate} \; \probability \, ( \bm{\force}_1 \cdot \bm{\velocity}_1 + \bm{\force}_2 \cdot \bm{\velocity}_2 ) \\
\label{eq:twowork}
\dot{\work} &= \int \d \bm{\coordinate} \, \dot{\energydensity} \, \probability + \int \d \bm{\coordinate} \; \probability \, ( \bm{\force}_1 \cdot \bm{\velocity}_1 + \bm{\force}_2 \cdot \bm{\velocity}_2 ) . 
\end{align} 
Using the Fokker-Planck Eq. \eqref{eq:twofokkerplanckequation}, we can write the heat current in terms of additive contributions,
\begin{align} \label{eq:twoheat}
\dot{\heat} = \sum_{i=1}^2 \dot{q}^{(i)}, \;  \dot{q}^{(i)} = -\friction_i \int \d \bm{\coordinate} \probability \left( \bm{\velocity}_i + \frac{1}{\invtemperature_i \mass_i} \D_{\bm{\velocity}_i} \ln \probability \right)  \bm{\velocity}_i .
\end{align}
Like in the single-particle case \eqref{eq:singleentropy}, the nonequilibrium system entropy is defined as
\begin{align} \label{eq:twoaverageentropy}
\entropy = - \int \d \bm{\coordinate} \, \probability \ln \probability , 
\end{align}
and the entropy balance is thus given by
\begin{align}  \label{eq:twoentropybalance}
\d_t \entropy = \sum_{i=1}^2 \invtemperature_i \, \dot{q}^{(i)} + \dot{\ep} , 
\end{align} 
where the non-negative entropy production rate
\begin{align} \label{eq:twoentropyproduction}
\dot{\ep} \!=\! \sum_{i=1}^2 \dot{\sigma}^{(i)}, \; \dot{\sigma}^{(i)} \!=\! \invtemperature_i \, \friction_i \!\! \int \!\! \d \bm{\coordinate} \, \probability \!\left( \bm{\velocity}_i \!+\! \frac{1}{\invtemperature_i \mass_i } \D_{\bm{\velocity}_i} \ln \probability \! \right)^2 \!\! \geq \! 0 ,
\end{align}
constitutes the second law of thermodynamics. In fact, Eq. \eqref{eq:twoentropyproduction} formulates a stronger statement: the additive contributions $\dot{\sigma}^{(i)}$ are separately non-negative. Yet, as will be shown further below, the additive contributions $\dot{\sigma}^{(i)}$ are, in general, not equal to the entropy production associated with the single particles.

%Using the definitions for the nonequilibrium free energy in Eq. \eqref{eq:singlefreenergy}, one can rewrite the work and heat current as follows
%\begin{align} \label{eq:twoheatworkcurrent} 
%\begin{aligned}
%\dot{\work} = \int \d \bm{\coordinate} \; \probability \, \dot{\freeenergydensity} , \quad 
%\dot{\heat} = d_t (\freeenergy + \invtemperature^{-1} \entropy) - \dot{\work}  ,
%\end{aligned}
%\end{align} 
%implying that the entropy production rate in Eq. \eqref{eq:twoentropyproduction} can be rewritten as 
%\begin{align}  \label{eq:twofreeenergysecondlaw} 
%\dot{\ep} = \invtemperature ( \dot{\work} - \d_t \freeenergy  ) \geq 0. 
%\end{align}

\section{Coarse graining}  
\label{sec:coarsegraining}
\subsection{Effective Dynamics}

We now shift our attention to the first particle alone. This formally amounts to integrating the Fokker-Plank Eq. \eqref{eq:twofokkerplanckequation} over the coordinates of the second particle $\bm{\coordinate}_2 ~=~ (\bm{\pos}_2,\bm{\velocity}_2)$ such that we obtain the marginalized probability distribution of particle one, $ \probability_1 \equiv \int \d\bm{\coordinate}_2 \, \probability$, that satisfies the following effective Fokker-Planck equation
\begin{align}  \label{eq:twocoarsegrainedfokkerplanckequation} 
\D_t \probability_1 
%&= - \D_{\bm{\pos}_1} (\bm{\velocity}_1 \probability_1) + \frac{ \friction_1 }{\invtemperature \mass_1^2} \D^2_{\bm{\velocity}_1} \probability_1  + \frac{1}{\mass_1} \D_{\bm{\velocity}_1} \big[ \big( \D_{\bm{\pos}_1} \potential_1(\bm{\pos}_1) - \bm{\force}(\bm{\coordinate}_1,t) - \bm{\force}_{1}(\bm{\coordinate}_1,t) + \friction_1 \bm{\velocity}_1 \big) \probability_1 \big] \\
&=  - \nabla_1 \cdot \currentvec_1 = - \nabla_1 \cdot \left( \bm{L}^{det}_1 + \bm{L}^{diss}_1 \right) \probability_1  ,
\end{align}
with $ \nabla_1 = ( \D_{\bm{\pos}_1} , \D_{\bm{\velocity}_1} )^\top$. The marginal probability current $\currentvec_1$ can be split into a deterministic part 
\begin{align}  \label{eq:twoprobabilitycurrentconservative}
\bm{L}^{det}_1 = \begin{pmatrix} \bm{\velocity}_1 \\ \frac{1}{\mass_1} \left[ - \D_{\bm{\pos}_1} \potential_1(\bm{\pos}_1,t) + \bm{\force}_1(\bm{\pos}_1,t) + \bm{\force}^{(1)}(\bm{\coordinate}_1,t) \right]  \end{pmatrix}   ,
\end{align} 
and a dissipative one
\begin{align}  \label{eq:twoprobabilitycurrentdissipative}
\bm{L}^{diss}_1 = 
\begin{pmatrix} 
0 \\ \frac{-\friction_1}{\mass_1^2} ( \mass_1 \bm{\velocity}_1 + \invtemperature_1^{-1} \D_{\bm{\velocity}_1} \ln \probability_1 ) 
\end{pmatrix} .
\end{align} 

By comparison with the exact single-particle Fokker-Planck Eq. \eqref{eq:singlefokkerplanck}, we note that the coarse-graining of the second particle encodes the interaction between the two particles in the effective and nonconservative force imposed on particle one 
\begin{align}  \label{eq:twonoconservativeforce}
\bm{\force}^{(1)}(\bm{\coordinate}_1,t) = - \int \d \bm{\coordinate}_2 \, \probability_{2|1}(\bm{\coordinate},t) \, \D_{\bm{\pos}_1} \potential^{int}_{12} (\bm{\pos}_1,\bm{\pos}_2,t).
\end{align}
We note that the evolution Eq. \eqref{eq:twocoarsegrainedfokkerplanckequation} is not closed since $\bm{\force}^{(1)}$ depends on $\probability_{2|1}$.
Thus, solving the effective Fokker-Planck Eq. \eqref{eq:twocoarsegrainedfokkerplanckequation} is as difficult as treating the full-Fokker-Planck Eq. \eqref{eq:twofokkerplanckequation}.

Moreover, for specific choices of the interaction potential, the first particle might be considered as an active Brownian particle. In this case, the velocity-dependent nonconservative force \eqref{eq:twonoconservativeforce} is interpreted as an additional energy inflow leading to active motion.
The latter is described effectively by negative dissipation in the direction of motion with velocity-dependent friction kernels. Some prominent models of active Brownian particles can, for instance, be found in Refs. \cite{schimansky2012epj,chaudhuri2013pre,chaudhuri2014pre,collet2019pre}.

\subsection{Effective Thermodynamics}
\label{sec:coarsegrainingthermo}
\subsubsection{Marginalization}

In the following, we attempt to formulate a consistent thermodynamic description for this reduced dynamics. Naively, it is tempting to use as an educated guess the single-particle expressions in Sec. \ref{sec:singletheory} for the reduced dynamics. In this case, the naive entropy balance reads
\begin{align}
\label{eq:twocoarsegrainedentropybalanceexclusivenaive} 
\d_t \entropy_{\bm{1}} = \invtemperature_1 \dot{q}^{(1)} + \dot{\ep}^{(1)} + \dot{\mutual}_F^{(1)}  \, , 
\end{align}
where we use the notation from Eq. \eqref{eq:singleinformationforce},
\begin{align} \label{eq:twocoarsegrainedinformationforce}
\dot{\mutual}_F^{(1)} \equiv \frac{1}{\mass_1}  \int \d \bm{\coordinate}_1 \, \probability_1 \, \D_{\bm{\velocity}_1} \cdot \bm{\force}^{(1)}  ,
\end{align}
and denote the single-particle Shannon entropy by
\begin{align} \label{eq:singleparticleshannonentropy}
\entropy_{\bm{1}} = - \int \d \bm{\coordinate}_1 \, \probability_1 \, \ln \probability_{1} ,
\end{align}
which implies for the non-negative effective entropy production rate
\begin{align}  \label{eq:twocoarsegrainedentropyproductionexclusive}
\dot{\ep}^{(1)} = \invtemperature_1 \friction_1 \! \int \d \bm{\coordinate}_1 \, \probability_1 \! \left( \bm{\velocity}_1 + \frac{1}{\invtemperature_1 \mass_1} \D_{\bm{\velocity}_1} \ln \probability_1 \right)^2 \geq 0  .
\end{align} 
For reasons that will become clear soon, we however define the effective entropy balance as follows,
\begin{align}
\label{eq:twocoarsegrainedentropybalanceexclusive} 
\d_t \entropy = \invtemperature_1 \dot{\heat}^{(1)} + \dot{\ep}^{(1)} + \dot{\mutual}_F^{(1)}  \, , 
\end{align}
where the effective heat
\begin{align} \label{eq:twocoarsegrainedheatexclusive}
\dot{\heat}^{(1)} = \dot{q}^{(1)} + \invtemperature_1^{-1} \entropy_{\bm{2}|\bm{1}} ,
\end{align}
is supplemented by the conditional Shannon entropy
\begin{align} \label{eq:conditionalshannonentropy}
\entropy_{\bm{2}|\bm{1}} = \entropy - \entropy_{\bm{1}} = - \int \d \bm{\coordinate}_1 \, \probability_1 \, \int \d \bm{\coordinate}_2 \, \probability_{2|1} \, \ln \probability_{2|1} .
\end{align}
The difference between the full \eqref{eq:twoheat} and effective \eqref{eq:twocoarsegrainedheatexclusive} heat current can be written as 
\begin{align}  \label{eq:twoheatdifferenceexclusive} 
\dot{\heat} - \dot{q}^{(1)} =  \dot{q}^{(2)} - \invtemperature_1^{-1} \entropy_{\bm{2}|\bm{1}} .
\end{align} 
Moreover, the difference between the full \eqref{eq:twoentropyproduction} and the effective entropy production rate \eqref{eq:twocoarsegrainedentropyproductionexclusive} is given by 
\begin{align}  \label{eq:twoentropyproductiondifferenceexclusive}
\dot{\ep} - \dot{\ep}^{(1)} = \int \d \bm{\coordinate}_1 \, \probability_1 \, \mathbb{\dot{\ep}}_1 ,
\end{align}
with the internal entropy production rate kernel
\begin{align}  \label{eq:twoentropyproductiondifferenceexclusivenotation}
\mathbb{\dot{\ep}}_1 = \mathbb{\dot{\ep}}_1' + \mathbb{\dot{\ep}}_1'' ,
\end{align}
that can be split in the following two non-negative contributions 
\begin{align}  \label{eq:twoentropyproductiondifferenceexclusivenotationone} 
\mathbb{\dot{\ep}}_1' &= \invtemperature_2 \, \friction_2 \! \int \! \d \bm{\coordinate}_2 \, \probability_{2|1} \! \left( \bm{\velocity}_2 \!+\!\!  \frac{1}{\invtemperature_2 \, \mass_2} \D_{\bm{\velocity}_2} \ln \probability_{2|1} \! \right)^2  \geq 0 \\ 
\mathbb{\dot{\ep}}_1'' &= \frac{\friction_1}{\invtemperature_1 \mass_1^2} \int \d \bm{\coordinate}_2 \, \probability_{2|1} \big( \D_{\bm{\velocity}_1} \ln \probability_{2|1} \big)^2 \geq 0 . \label{eq:twoentropyproductiondifferenceexclusivenotationtwo}
\end{align}
The first contribution $\mathbb{\dot{\ep}}_1'$ is the entropy production rate of the second particle if the coordinates of the first one are fixed, see Eq. \eqref{eq:twoentropyproduction}. Conversely, the second contribution $\mathbb{\dot{\ep}}_1''$ can be viewed as a contribution to the entropy production rate due to the correlation of the particles as we will see in Eq. \eqref{eq:mutualinformationderivativeone}.

An equivalent decomposition to Eq. \eqref{eq:twoentropyproductiondifferenceexclusivenotation} for Markovian master equations was found in Ref. \cite{esposito2012pre}.
From the last two equations we deduce that the effective entropy production (rate) always underestimates the physical one
\begin{align} \label{eq:entropyproductionraterelationpartial} 
\dot{\ep} \geq \dot{\ep}^{(1)} .
\end{align}
It is important to note that at this general level it is impossible to fully capture the full thermodynamics solely in terms of properties of the reduced dynamics. The missing contributions require knowledge about the conditional probability $\probability_{2|1}$.

\subsubsection{Bipartite System}

A second approach to formulate an effective thermodynamics is provided by a bipartite system where the two-particle system is split into two single-particle subsystems. The effective entropic expressions in both subsystems are defined in the same formal way as one would for a single particle. Subsequently, the sum of the effective entropy balances in both subsystems is compared with the full one of the two-particle system in order to identify the so-called information flows exchanged between the subsystems.

Physically, a bipartite system provides a simple and convenient representation of a Maxwell’s demon since the thermodynamic cost of the latter becomes fully accessible \cite{esposito2014prx,horowitz2015jstatmech,hartrich2014jsm}.
Mathematically, the bipartite structure identifies the non-additive contributions of the full thermodynamic quantities for the two particles.
We first note that the additive contributions to the two-particle heat current \eqref{eq:twoheat} can be rewritten in terms of marginalized probabilities only as follows 
\begin{align} \label{eq:twoheatrewritten} 
\dot{q}^{(i)} = -\friction_i \int \d \bm{\coordinate}_i \probability_i \left( \bm{\velocity}_i + \frac{1}{\invtemperature_i \mass_i} \D_{\bm{\velocity}_i} \ln \probability_i \right)  \bm{\velocity}_i ,
\end{align}
where the marginal probability $\probability_2$ is obtained analogously as $\probability_1$, that is by marginalizing the two-point probability $\probability$ over $\bm{\coordinate}_1$.
Using the last Eq. along with Eqs. \eqref{eq:singleheat} and \eqref{eq:twoheat}, we see that the following relation holds,
\begin{align} \label{eq:twoheatadditivehamiltonian} 
\dot{q}^{(i)} = \int \d \bm{\coordinate}_i \, \energydensity_i \, \dot{\probability_i} - \int \d \bm{\coordinate}_i \; \probability_i \; \bm{\velocity}_i \cdot \left( \bm{\force}_i + \bm{\force}^{(i)} \right) ,
\end{align}
with the nonconservative force $\bm{\force}^{(2)}$ 
\begin{align}  \label{eq:twonoconservativeforcetwo}
\bm{\force}^{(2)}(\bm{\coordinate}_2,t) = - \int \d \bm{\coordinate}_{1} \, \probability_{1|2}(\bm{\coordinate},t) \; \D_{\bm{\pos}_2} \potential^{int}_{12} (\bm{\pos}_1,\bm{\pos}_2,t) .
\end{align}

Conversely, the additive contributions $\dot{\sigma}^{(i)}$ to the entropy production rate in Eq. \eqref{eq:twoentropyproductionrewritten} can not be represented by marginal distributions only. Therefore, the entropy-balance equations for the subsystems of the bipartite system can not be expressed in terms of its associated degrees of freedom only. We proceed by deriving the non-additive contribution to the entropy and identifying it as the information flow.

To this end, we first define the relative entropy (or Kulback-Leibler divergence) as a statistical measure of the distance between the distributions $\probability$ and $\probability_1 \probability_2$ as follows
\begin{align}  \label{eq:mutualinformation}
\mutual = D[\probability \, || \, \probability_1 \probability_2] =  \int \d \bm{\coordinate} \, \probability \ln \frac{\probability}{\probability_1 \probability_2} \geq 0 , 
\end{align} 
whose non-negativity readily follows from the inequality $\ln \probability \leq \probability-1$. 
From Eqs. \eqref{eq:singleentropy} and \eqref{eq:twoaverageentropy} follows that the relative entropy is the non-additive part of the two-particle system entropy, \idest
\begin{align}  \label{eq:mutualinformationtwo}
\mutual = \entropy_{\bm{1}} + \entropy_{\bm{2}} - \entropy  .
\end{align} 
Physically, this quantity corresponds to the mutual information that is a measure of correlations that quantifies how much one system knows about the other. If $\mutual$ is large, the two systems are highly correlated, whereas small values of $\mutual$ imply that the two systems know little about each other. The time-derivative of the mutual information
\begin{align}  \label{eq:mutualinformationderivative}
\d_t \mutual = \dot{\mutual}^{(2 \rightarrow 1)} + \dot{\mutual}^{(1 \rightarrow 2)}  ,
\end{align}
can be split into two directional information flows 
\begin{align} \label{eq:mutualinformationderivativeone}
\dot{\mutual}^{(2 \rightarrow 1)} &= \int \d \bm{\coordinate}_1 \probability_1 \; 
\left( \frac{1}{\mass_1} \D_{\bm{\velocity}_1} \cdot \bm{\force}^{(1)} - \mathbb{\dot{\ep}}_1'' \right) \\
\dot{\mutual}^{(1 \rightarrow 2)} &= \int \d \bm{\coordinate}_2 \probability_2 \; 
\left( \frac{1}{\mass_2} \D_{\bm{\velocity}_2} \cdot \bm{\force}^{(2)} - \mathbb{\dot{\ep}}_2'' \right) , 
\label{eq:mutualinformationderivativetwo}
\end{align}
where we used Eqs. \eqref{eq:twonoconservativeforce} and \eqref{eq:twoentropyproductiondifferenceexclusivenotationtwo} in the first equation.
In the second equation we used Eq. \eqref{eq:twonoconservativeforcetwo} and introduced the integral kernel specifying the difference between the full and the effective entropy production rate for the second particle, 
\begin{align}  \label{eq:twoentropyproductiondifferenceexclusivetwo}
\dot{\ep} - \dot{\ep}^{(2)} = \int \d \bm{\coordinate}_{2} \, \probability_{2} \, \mathbb{\dot{\ep}}_{2} = \int \d \bm{\coordinate}_{2} \, \probability_{2} ( \mathbb{\dot{\ep}}_{2}' + \mathbb{\dot{\ep}}_{2}'') , 
\end{align}
with
\begin{align} \label{eq:twoentropyproductiondifferenceexclusivenotationonetwo}
\mathbb{\dot{\ep}}_{2}' &= \invtemperature_1 \, \friction_1  \int  \d \bm{\coordinate}_{1} \, \probability_{1|2} \left( \bm{\velocity}_1 + \frac{1}{\invtemperature_1 \, \mass_1} \D_{\bm{\velocity}_1} \ln \probability_{1|2} \right)^2 \geq 0 \\ 
\mathbb{\dot{\ep}}_{2}'' &= \frac{\friction_2}{\invtemperature_2 \mass_2^2} \int \d \bm{\coordinate}_{1} \, \probability_{1|2} \big( \D_{\bm{\velocity}_2} \ln \probability_{1|2} \big)^2  \geq 0 .
\label{eq:twoentropyproductiondifferenceexclusivenotationtwotwo}
\end{align} 
The directional information flows can be interpreted as follows: When $\dot{\mutual}^{(i \rightarrow j)} > 0$, the dynamics of particle $j$ increases the mutual information and thus the correlations between the two particles. In other words, $j$ is learning about $i$ and vice versa.
Conversely, $\dot{\mutual}^{(i \rightarrow j)} < 0$ corresponds to decreasing correlations between the two particles due to the evolution of particle $j$, which can be interpreted as either information erasure or the conversion of information into energy \cite{esposito2014prx}.
We furthermore point out that a positive directional information flow indicates that its force contribution
\begin{align} \label{eq:mutualinformationforce} 
\dot{\mutual}^{(i \rightarrow j)}_F &\equiv \frac{1}{\mass_j} \int \d  \bm{\coordinate}_j \, \probability_j \; \D_{\bm{\velocity}_j} \cdot \bm{\force}^{(j)} ,
\intertext{dominates its entropic part}
\dot{\mutual}^{(i \rightarrow j)}_{\entropy} &\equiv - \int \d \bm{\coordinate}_j \, \probability_j \; \mathbb{\dot{\ep}}_j'' 
\label{eq:mutualinformationentropy}  ,
\end{align}
since the latter is non-positive according to Eq. \eqref{eq:twoentropyproductiondifferenceexclusivenotationtwo}.
Various other interpretations of these mutual information flows have been discussed in the literature 
\cite{schreiber2000measuring, allahverdyan2009thermodynamic, liang2005information, majda2007information, barato2013pre, seifert2017prl}.

An inspection of Eq. \eqref{eq:twocoarsegrainedinformationforce} reveals that the force contribution of the information flow, $\dot{\mutual}^{(i \rightarrow j)}_F$, is the additional term that enters in the effective entropy balance due to the velocity-dependent nonconservative force $\bm{\force}^{(j)}$,
\begin{align}
\d_t \entropy_{\bm{j}} = \invtemperature_j \dot{q}^{(j)} + \dot{\ep}^{(j)} + \dot{\mutual}^{(i \rightarrow j)}_F . \label{eq:twocoarsegrainedentropybalancemutualinformationtwo} 
\end{align}
Using Eq. \eqref{eq:mutualinformationentropy}, we furthermore find that the difference between the effective \eqref{eq:twocoarsegrainedentropyproductionexclusive} and the additive contribution to the two-particle entropy production rate \eqref{eq:twoentropyproduction}
corresponds to the entropic part of the information flow,
\begin{align}   \label{eq:twoentropyproductionrewritten}
\dot{\mutual}^{(i \rightarrow j)}_{\entropy} = \dot{\ep}^{(j)} - \dot{\sigma}^{(j)} .
\end{align} 
The last two equations stipulate the following effective entropy balance equation for particle $j$,
\begin{align}  \label{eq:twocoarsegrainedentropybalancemutualinformation} 
\d_t \entropy_{\bm{j}} = \invtemperature_j \dot{q}^{(j)} + \dot{\sigma}^{(j)} + \dot{\mutual}^{(i \rightarrow j) } .
\end{align} 
It is important to note that Eq. \eqref{eq:twocoarsegrainedentropybalancemutualinformation} states that the directional information flows are the non-additive quantities entering in the effective entropy balance. We emphasize that Eq. \eqref{eq:twocoarsegrainedentropybalancemutualinformation} is the underdamped Fokker-Planck analog of the result found for master equations in Ref. \cite{esposito2014prx}.
Moreover, using Eqs. \eqref{eq:twoentropyproductiondifferenceexclusivenotationone} and \eqref{eq:twoentropyproductionrewritten}, it holds that 
\begin{align}
\int \d \bm{\coordinate}_i \, \probability_i \, \dot{\mathbb{\ep}}_i' = \dot{\sigma}^{(j)}  = \dot{\ep}^{(j)} - \dot{\mutual}^{(i \rightarrow j)}_{\entropy} ,
\end{align}
which because of Eq. \eqref{eq:twoentropyproduction} implies that
\begin{align}  \label{eq:entropyproductioncomparison}
\dot{\ep} = \dot{\ep}^{(1)} + \dot{\ep}^{(2)} - \dot{\mutual}^{(2 \rightarrow 1)}_{\entropy} - \dot{\mutual}^{(1 \rightarrow 2)}_{\entropy},
\end{align} 
An identical result for bipartite master equations was found in Ref. \cite{esposito2014jstatmech} and recently for the more general case of systems undergoing a quantum dynamics formulated in terms of a density matrix, where the generator is additive with respect to the reservoirs \citep{esposito2019prl}.

\subsubsection{Hamiltonian of Mean Force}

Finally, we present a third approach to define an effective thermodynamics for the reduced dynamics of particle $\bm{1}$ in Fig. \ref{fig:modelschematics}, where we set $\invtemperature_{1,2} = \invtemperature$ and $\bm{\force}_{2}=0$.
For reasons that will become clear soon, we furthermore consider an explicitly time-independent bare Hamiltonian of the second particle $ \D_t \energydensity_2 = \D_t \potential_2 = 0$. As we will see, for this approach only a specific class of initial conditions can be considered.

The key concept is the so-called Hamiltonian of mean force, originally utilized in equilibrium thermostatics \cite{kirkwood1935jcp}, which defines an effective energy for particle $\bm{1}$ that accounts for the strong coupling \cite{jarzynski2017prx} to the second particle $\bm{2}$.  Using it, this approach attempts to overcome the problem identified in the context of Eq. \eqref{eq:twocoarsegrainedentropybalanceexclusive} that there is \emph{a priori} no systematic way to embed the global energetics into the reduced dynamics. 

The marginal of the global (Gibbs) equilibrium distribution over the second particle can be expressed as
\begin{align} \label{eq:marginal1HMF}
\probability_1^{hmf} \!=\!\! \int \!\! \d \bm{\coordinate}_2 \, \probability^{eq} \!=\!\! \int \!\! \d \bm{\coordinate}_2 \, \euler^{-\invtemperature (\energydensity - \freeenergy^{eq} ) } \!=\! \euler^{- \invtemperature (\hamiltonian^{hmf} - \freeenergy_{hmf}^{eq} ) }  ,
\end{align}
where we introduced the effective free energy $\freeenergy_{hmf}^{eq}$ of particle one which is defined as the difference between the full equilibrium free energy
\begin{align} \label{eq:equilibriumfreeenergy}
\freeenergy^{eq} = - \frac{1}{\invtemperature} \, \ln \int \d \bm{\coordinate} \; \euler^{-\invtemperature \energydensity }  ,
\end{align}
and that of the second particle
\begin{align} \label{eq:equilibriumfreeenergytwo}
\freeenergy^{eq}_{\bm{2}} = - \frac{1}{\invtemperature} \, \ln \int \d \bm{\coordinate}_2 \; \euler^{-\invtemperature \energydensity_2 }  ,
\end{align}
that is $\freeenergy_{hmf}^{eq} = \freeenergy^{eq} - \freeenergy_{\bm{2}}^{eq}$.
Consequently the HMF is defined as
\begin{align} \label{eq:hmfdefinition} 
\hamiltonian^{hmf} \equiv \energydensity_1 - \invtemperature^{-1} \, \ln \, \langle \euler^{- \invtemperature \potential^{int}_{12} } \rangle^{eq}_{\bm{2}} .
\end{align}
We denote by $\langle \cdot \rangle^{eq}_{\bm{2}}$ and $\langle \cdot \rangle^{eq}$ an ensemble average over the equilibrium distribution of particle two, $\probability_2^{eq} ~=~ \exp[-\invtemperature (\energydensity_2 - \freeenergy^{eq}_{\bm{2}})]$, and over the global equilibrium distribution, respectively.

The conditional equilibrium distribution $\probability_{2|1}^{eq}$ is obtained by dividing the global (Gibbs) equilibrium distribution by the marginal one in Eq. \eqref{eq:marginal1HMF}
\begin{align} \label{eq:equilibriumconditionalprobability}
\probability_{2|1}^{eq} = \frac{\probability^{eq}}{\probability_1^{hmf}} = \euler^{- \invtemperature \left( \energydensity - \freeenergy_{\bm{2}|1}^{eq} \right) } ,
\end{align}
where the free-energy landscape of particle one for a conditionally equilibrated particle two is 
\begin{align} \label{eq:freeenergylandscape} 
\freeenergy_{\bm{2}|1}^{eq} = \energydensity_1 - \invtemperature^{-1} \, \ln \, \langle \euler^{- \invtemperature \potential^{int}_{12} } \rangle^{eq}_{\bm{2}} + \freeenergy_{\bm{2}}^{eq} = \hamiltonian^{hmf} + \freeenergy_{\bm{2}}^{eq} .
\end{align}
It is noteworthy that $\freeenergy_{\bm{2}|1}^{eq}$ is parametrically time-dependent, whereas $\freeenergy_{\bm{2}}^{eq}$ has no time-dependence due to the choice of a time-independent Hamiltonian $\energydensity_2$.
Eq. \eqref{eq:freeenergylandscape} shows that up to $\freeenergy_{\bm{2}}^{eq}$, the HMF is equal to the free energy that the locally equilibrated second particle generates for given coordinates of the first particle.

Furthermore, we note the standard equilibrium identities
\begin{alignat}{2}
\freeenergy^{eq}_{\bm{2}|1} &= \energy^{eq}_{\bm{2}|1} -\! \invtemperature^{-1} \entropy^{eq}_{\bm{2}|1} ,\label{eq:conditionalfreeenergykernelfast} \\ 
\energy^{eq}_{\bm{2}|1} &= \D_{\invtemperature} (\invtemperature \freeenergy_{\bm{2}|1}^{eq} ) 
= \int \d \bm{\coordinate}_2 \, \probability_{2|1}^{eq} \, \energydensity \label{eq:conditionalenergykernelfast} \\ 
\entropy^{eq}_{\bm{2}|\bm{1}} &= \invtemperature^2 \D_{\invtemperature} \freeenergy_{\bm{2}|1}^{eq} = - \int \d \bm{\coordinate}_2 \, \probability_{2|1}^{eq} \, \ln \probability_{2|1}^{eq} \label{eq:conditionalshannonentropykernelfast} ,
\end{alignat}
which, using Eq. \eqref{eq:hmfdefinition}, can be rewritten as
\begin{align}
\energy^{eq}_{\bm{2}|1} &= \D_{\invtemperature} \big[ \invtemperature \big( \hamiltonian^{hmf} + \freeenergy_{\bm{2}}^{eq} \big) \big] 
\label{eq:conditionalenergykernelfastrewritten} \\ 
\entropy^{eq}_{\bm{2}|\bm{1}} &= \invtemperature^2 \, \D_{\invtemperature} \big( \hamiltonian^{hmf} + \freeenergy_{\bm{2}}^{eq} \big)  \label{eq:conditionalshannonentropykernelfastrewritten} .
\end{align}

Inspired by \cite{seifert2016prl}, we employ the HMF \eqref{eq:hmfdefinition} and its derived quantities in Eqs. \eqref{eq:conditionalenergykernelfastrewritten} and \eqref{eq:conditionalshannonentropykernelfastrewritten} and average them over arbitrary \emph{nonequilibrium} probabilities for particle one, \idest
\begin{align} \label{eq:hmffirstlaw}
\energy^{hmf} (t) = \langle \D_{\invtemperature} ( \invtemperature \, \hamiltonian^{hmf} ) \rangle (t) ,
\end{align}
and 
\begin{align} \label{eq:hmfentropy}
\entropy^{hmf}(t) &\equiv \entropy_{\bm{1}}(t) + \invtemperature^{2} \langle \D_{\invtemperature} \, \hamiltonian^{hmf} \rangle (t) ,
\end{align}
where $\langle \cdot \rangle(t)$ refers to an ensemble average over a generic \emph{nonequilibrium} distribution $\probability(t)$. We note that the definition of the entropy \eqref{eq:hmfentropy} also includes the single-particle Shannon entropy of particle one in addition to the contribution that stems from the HMF.
Choosing a definition of work that coincides with the global one \eqref{eq:twowork},
\begin{align} \label{eq:hmfwork}
\work^{hmf}(t) \equiv \int\limits_0^t \d t' \, \Big[ \, \langle  \dot{\energydensity} \rangle (t') 
+ \Big( \int \d \bm{\coordinate}_1 \; \probability_1 \; \bm{\velocity}_1 \cdot \bm{\force}_{1} \Big)(t') \Big] , 
\end{align}
the first law of thermodynamics imposes the following definition for heat 
\begin{align}  \label{eq:hmfheatrewritten}
\heat^{hmf}(t) \! = \! - \work(t) \!+\! \langle \D_{\invtemperature} ( \invtemperature \, \hamiltonian^{hmf} ) \rangle (t) \!-\! \langle \D_{\invtemperature} ( \invtemperature \, \hamiltonian^{hmf} ) \rangle (0) .
\end{align}
Defining the nonequilibrium free energy to be of the same form as in the standard equilibrium case \eqref{eq:conditionalfreeenergykernelfast},
\begin{align} \label{eq:hmffreeenergy} 
\freeenergy^{hmf}(t) = \energy^{hmf}(t) - \frac{\entropy^{hmf}(t)}{\invtemperature} = \langle \hamiltonian^{hmf} \rangle (t) - \frac{\entropy_{\bm{1}}(t)}{\invtemperature} ,
\end{align}
we can rewrite the entropy balance
\begin{align}  \label{eq:hmfsecondlaw}
\Delta \entropy^{hmf}(t) = \invtemperature \heat^{hmf}(t) + \ep^{hmf}(t) , 
\end{align}
in the form of a second law of thermodynamics as follows 
\begin{align} \label{eq:hmffreeenergysecondlaw} 
\ep^{hmf}(t) = \invtemperature \big[ \work(t) - \Delta \freeenergy^{hmf}(t) \big] \geq 0. 
\end{align}
In order to prove the non-negativity of this definition for the entropy production \cite{seifert2016prl,strasberg2017pre}, an initial condition of the form
\begin{align} \label{eq:hmfconditionalprobability}
\probability(0) = \probability_{1}(0) \, \probability_{2|1}^{eq} = \probability_{1}(0) \, \euler^{- \invtemperature \left( \energydensity - \hamiltonian^{hmf} - \freeenergy_{\bm{2}}^{eq} \right) }  ,
\end{align}
is required.
Indeed, using Eqs. \eqref{eq:singlefreenergy} and \eqref{eq:hmffreeenergysecondlaw}, we have
\begin{align}
\ep^{hmf}(t) - \ep(t) = \invtemperature \left( \Delta \freeenergy - \Delta \freeenergy^{hmf}(t) \right) .
\end{align}
Due to the special choice for the initial condition \eqref{eq:hmfconditionalprobability}, Eqs. \eqref{eq:conditionalenergykernelfastrewritten} and \eqref{eq:conditionalshannonentropykernelfastrewritten} are valid at $t=0$ so that
\begin{align} \label{eq:hmffreeenergytss} 
\freeenergy(0) - \freeenergy^{hmf}(0) = \freeenergy_{\bm{2}}^{eq} . 
\end{align}
At later times, Eqs. \eqref{eq:conditionalenergykernelfastrewritten} and \eqref{eq:conditionalshannonentropykernelfastrewritten} are no longer valid and we need to resort to the definitions \eqref{eq:hmfentropy} and \eqref{eq:hmffirstlaw} to obtain
\begin{align}
\freeenergy(t) \!-\! \freeenergy^{hmf}(t) \!=\! \langle \energydensity \rangle(t) \!-\! \langle \hamiltonian^{hmf} \! \rangle(t) \!+\! \invtemperature^{-1}  [ \entropy_{\bm{1}}(t) \!-\! \entropy(t) ] .
\end{align}
Since the HMF can also be expressed as
\begin{align}
\langle \hamiltonian^{hmf} \rangle(t) = \langle \energydensity \rangle(t) + \invtemperature^{-1} \langle \ln \probability_{2|1}^{eq} \rangle - \freeenergy_{\bm{2}}^{eq} ,
\end{align}
we have
\begin{align} 
\freeenergy(t) - \freeenergy^{hmf}(t) = \freeenergy_{\bm{2}}^{eq} 
+ \invtemperature^{-1} \Big\langle \ln \frac{\probability(t)}{\probability_{2|1}^{eq} \, \probability_{1}(t)} \Big\rangle , 
\end{align}
and finally arrive at
\begin{align} \label{eq:entropyproductionraterelationpartialtwo}
\ep^{hmf}(t) - \ep(t) = D[\probability(t) \, || \, \probability_{2|1}^{eq} \probability_{1}(t) ]  \geq 0 .
\end{align}
Thus, the entropy production based on the HMF always overestimates the global two-particle entropy production
which, because of Eq. \eqref{eq:twoentropyproduction}, proves the inequality in Eq. \eqref{eq:hmffreeenergysecondlaw}. Furthermore, with Eq. \eqref{eq:entropyproductionraterelationpartial} we obtain the following hierarchies of inequalities
\begin{align} \label{eq:entropyproductionraterelationfull}
\ep^{hmf}(t) \geq \ep(t) \geq \ep^{(1)}(t) ,
\end{align}
where the equality signs hold in the limit of time-scale separation, as will be shown further below. The last equation is the Fokker-Planck analog of the result found for master equations in Ref. \cite{strasberg2017pre}. This reference also identifies the conditions under which the \emph{rate} of the entropy production \eqref{eq:hmffreeenergysecondlaw} is non-negative.

\subsection{Limiting Cases}  
\label{sec:twospecialcases}

As already pointed out above, the effective Fokker-Planck Eq. \eqref{eq:twocoarsegrainedfokkerplanckequation} is, in general, not closed because of the dependence on the conditional probability $\probability_{2|1}$. With the results of the preceding section at hand, we now study the three different coarse-graining schemes for two limiting cases in which the effective Fokker-Planck equation becomes closed and thus analytically tractable.

\subsubsection{Fast-Dynamics Limit: The Heat Reservoir}
\label{sec:twofastparticle}

First, we assume a time-scale separation between the stochastic dynamics of the two particles where particle two evolves much faster than particle one. Hence for fixed coordinates of the first particle, the second generically relaxes towards a nonequilibrium steady state and the stationary conditional probability $\probability_{2|1}^{tss}$ can be determined by solving the fast dynamics for fixed $\bm{\coordinate}_1$. As a consequence, the effective Fokker-Planck Eq. \eqref{eq:twocoarsegrainedfokkerplanckequation} becomes closed and the effective thermodynamics follows from replacing $\probability_{2|1}$ by $\probability_{2|1}^{tss}$ in all expressions in Sec. \ref{sec:coarsegrainingthermo}. However, this effective thermodynamics naturally does not match with the full one, as we would neglect hidden degrees of freedom that are out-of-equilibrium.

The latter equilibrate only if $\bm{\force}_2 =0$ and $\invtemperature_{1,2}=\invtemperature$, that is when the second particle instantaneously equilibrates with respect to each value of the slow coordinates of particle one.
Then, the conditional probability is given at any time by the Gibbs distribution \cite{esposito2012pre} 
\begin{align}  \label{eq:conditionalprobabilityfastparticle} 
 \probability_{2|1}^{tss}(\bm{\pos}_1,\bm{\coordinate}_2) \equiv \probability_{2|1}^{eq}(\bm{\pos}_1,\bm{\coordinate}_2) 
= \euler^{- \invtemperature \left( \energydensity - \freeenergy_{\bm{2}|1}^{eq} \right) } .
\end{align}  
As a result, the effective force $\bm{\force}^{(1)}$ in Eq. \eqref{eq:twonoconservativeforce}, becomes a velocity-independent force that derives from an effective potential so that 
\begin{align}  \label{eq:equilibriumconservativeforce}
\left. \big[ - \D_{\bm{\pos}_1} \potential_1 + \bm{\force}^{{(1)}} \big] \right|_{tss}
% = -\D_{\bm{\pos}_1} \! \left[ \freeenergy_{\bm{2}|1}^{eq} - \potential_1 \right] 
= -\D_{\bm{\pos}_1} \freeenergy_{\bm{2}|1}^{eq}  ,
\end{align}
where the notation $\left. Z \right|_{tss}$ corresponds to the conditional probability $\probability_{2|1}$ in the expression $Z$ being substituted by the equilibrium one in Eq. \eqref{eq:conditionalprobabilityfastparticle}.
Hence in the limit of TSS and local equilibrium, the particle is subjected to the effective potential given by the free-energy landscape of the first particle, $ \freeenergy_{\bm{2}|1}^{eq} $. \\

\paragraph*{Marginalization.}
Substituting Eq. \eqref{eq:conditionalprobabilityfastparticle} into Eq. \eqref{eq:conditionalshannonentropy} and accounting for probability conservation, we get
\begin{align} \label{eq:conditionalshannonentropyfast}
\frac{\left. \d_t \entropy_{\bm{2}|\bm{1}} \right|_{tss} }{\invtemperature}
&= \int \d \bm{\coordinate} \,  \dot{\probability}_1(t) \, \probability_{2|1}^{eq} \left( \energydensity - \freeenergy_{\bm{2}|1}^{eq} \right) \\
&= \int \d \bm{\coordinate} \,  \dot{\probability}_1(t) \, \probability_{2|1}^{eq} \, \energydensity - \int \d \bm{\coordinate}_1 \, \dot{\probability}_1(t) \,  \freeenergy_{\bm{2}|1}^{eq} .
\end{align}
With Eqs. \eqref{eq:twoheatadditivehamiltonian} and \eqref{eq:freeenergylandscape}, we note the relation
\begin{align} 
\left. \dot{q}^{(1)} \right|_{tss} = \int \d \bm{\coordinate}_1 \,  \dot{\probability}_1(t) \,  \freeenergy_{\bm{2}|1}^{eq} ,
\end{align}
from which along with Eq. \eqref{eq:twocoarsegrainedheatexclusive} follows that
\begin{align} \label{eq:heatcurrentcoincidencefast}
\left. \dot{\heat}^{(1)} \right|_{tss} = \left. \dot{q}^{(1)} \right|_{tss} + \left. \invtemperature^{-1} \d_t \entropy_{\bm{2}|\bm{1}} \right|_{tss} = \left. \dot{\heat} \right|_{tss} ,
\end{align}
hence clarifying why the effective heat \eqref{eq:twocoarsegrainedheatexclusive} was defined to contain the conditional Shannon entropy.

We have therefore proven that in the limit of TSS the effective \eqref{eq:twocoarsegrainedheatexclusive} and the global heat current \eqref{eq:twoheatdefintion} coincide and the first law of thermodynamics remains formally the same as in Eq. \eqref{eq:twofirstlaw},
\begin{align} \label{eq:firstlawfastparticle}
\left. \d_t \energy \right|_{tss} = \left. \dot{\heat}^{(1)} \right|_{tss} + \left. \dot{\work} \right|_{tss} = \left. \dot{\heat} \right|_{tss} + \left. \dot{\work} \right|_{tss} . 
\end{align}
Furthermore, in the limit of TSS, the time-dependence of all quantities stems only from the dynamics of particle one and the parametric time-dependence of the Hamiltonian. 
Equation \eqref{eq:heatcurrentcoincidencefast} proves that the second law of thermodynamics formally also remains the same as in Eq. \eqref{eq:twoentropyproduction},
\begin{align}  \label{eq:secondlawfastparticle}
\left. \dot{\ep}^{(1)} \right|_{tss} &= \left. \d_t \entropy \right|_{tss} - \left. \invtemperature \dot{\heat}^{(1)} \right|_{tss} \\
&= \left. \d_t \entropy \right|_{tss} - \left. \invtemperature \dot{\heat} \right|_{tss} = \left. \dot{\ep} \right|_{tss} \geq 0 .
\end{align}
Hence in the limit of TSS, the full thermodynamics of the two particles can be described solely by the reduced dynamics of a single particle that is subjected to the potential $ \freeenergy_{\bm{2}|1}^{eq} $.
Physically, the second particle can be viewed as being part of the heat reservoir the first particle is coupled to. \\

\paragraph*{Bipartite System.}
Furthermore, substituting \eqref{eq:conditionalprobabilityfastparticle} into Eqs. \eqref{eq:mutualinformationderivativeone}, \eqref{eq:mutualinformationforce} and \eqref{eq:mutualinformationentropy}, gives a vanishing directional information flow from the fast to the slow particle,
\begin{align} \label{eq:informationflowfastparticle}
\left. \dot{\mutual}^{(2 \rightarrow 1)}_F \right|_{tss} = \left. \dot{\mutual}^{(2 \rightarrow 1)}_S \right|_{tss} = \left. \dot{\mutual}^{(2 \rightarrow 1)} \right|_{tss} = 0. 
\end{align}
This means that in the limit of TSS the information flow is completely asymmetric, $ \left. \d_t \, \mutual \right|_{tss} = \left. \d_t \, \mutual^{(1 \rightarrow 2)} \right|_{tss} $.
From the last equation follows that the additive and effective entropy production rate \eqref{eq:twocoarsegrainedentropyproductionexclusive} agrees with the global one \eqref{eq:twoentropyproduction},
\begin{align}
\left. \dot{\sigma}^{(1)} \right|_{tss} = \left. \dot{\ep}^{(1)} \right|_{tss}= \left. \dot{\ep} \right|_{tss} ,
\end{align}
which in turn implies that $\left. \dot{\sigma}^{(2)} \right|_{tss} = 0$.
Though, there is a mismatch between the effective entropy balance of the slow particle \eqref{eq:twocoarsegrainedentropybalancemutualinformation} and the full entropy balance \eqref{eq:twoentropybalance} given by the conditional Shannon entropy, 
\begin{align} \label{eq:twocoarsegrainedentropybalancemutualinformationfast} 
\left. \d_t \entropy_{\bm{1}} \right|_{tss} = \left. \invtemperature \dot{q}^{(1)} \right|_{tss} + \left. \dot{\sigma}^{(1)} \right|_{tss} 
%= \invtemperature \dot{\heat} - \d_t \entropy_{\bm{2}|\bm{1}} + \dot{\ep} 
= \left. \d_t \entropy \right|_{tss} - \left. \d_t \entropy_{\bm{2}|\bm{1}} \right|_{tss} .
\end{align}
Moreover, the effective entropy balance of the second particle reads
\begin{align} \label{eq:twocoarsegrainedentropybalancemutualinformationfasttwo}
\left. \d_t \entropy_{\bm{2}} \right|_{tss} = \invtemperature \, \left. \d_t q^{(2)} \right|_{tss} + \left. \d_t \mutual^{(1 \rightarrow 2)} \right|_{tss} ,
\end{align}
that can be rewritten as 
\begin{align}  \label{eq:twocoarsegrainedentropybalancemutualinformationfasttworewritten} 
\left. \d_t \entropy_{\bm{2}} \right|_{tss} - \left. \d_t \entropy_{\bm{2}|\bm{1}} \right|_{tss} = \left. \d_t \mutual^{(1 \rightarrow 2)} \right|_{tss} .
\end{align}
Equation \eqref{eq:twocoarsegrainedentropybalancemutualinformationfasttworewritten} stipulates that the information flow $\left. \d_t \mutual \right|_{tss} ~=~ \left. \d_t \mutual^{(1 \rightarrow 2)} \right|_{tss} $ from the slow to the fast particle does, in general, not vanish. This is physically plausible since the particles are still correlated. The information flow $\left. \d_t \mutual^{(1 \rightarrow 2)} \right|_{tss} $ reflects time-varying correlations between the two particles due to the change of the probability distribution of both out-of-equilibrium particles. Consequently, the information flow is zero for a global equilibrium state characterized by $\probability^{eq} = \probability_{2|1}^{eq} \, \probability_1^{eq} $. \\

\paragraph*{Hamiltonian of Mean Force.}
We now turn to the HMF formalism in the limit of TSS and local equilibrium, $\invtemperature_{1,2}=\invtemperature$ and $\bm{\force}_2=0$. Further, as done above in the introduction of the HMF formalism, we assume that the bare Hamiltonian of the second particle is time-independent, $ \D_t \, \energydensity_2 = 0$.
Because of Eq. \eqref{eq:conditionalprobabilityfastparticle}, the requirement of an initial  equilibrium conditional probability distribution \eqref{eq:hmfconditionalprobability} is fulfilled at all times $t$. 
Hence Eqs. \eqref{eq:conditionalenergykernelfastrewritten} and \eqref{eq:conditionalshannonentropykernelfastrewritten} are valid at any time $t$ and a comparison with Eqs. \eqref{eq:hmffirstlaw} and \eqref{eq:hmfentropy}, respectively, shows that
\begin{align} \label{eq:hmfenergytss} 
\left. \energy^{hmf}(t) \right|_{tss} &= \left. \energy (t) \right|_{tss} - \D_{\invtemperature} (\invtemperature \, \freeenergy^{eq}_{\bm{2}}) \\
\left. \entropy^{hmf} (t) \right|_{tss} &= \left. \entropy(t) \right|_{tss} - \invtemperature^2 \, \D_{\invtemperature} \, \freeenergy^{eq}_{\bm{2}}  \label{eq:hmfentropytss} .
\end{align}
This explains the choice of a time-independent Hamiltonian $\energydensity_2$, since in this case $\freeenergy_{\bm{2}}^{eq}$ has no time-dependence. As a result, the HMF definitions of the corresponding \emph{currents} coincide with the global ones,
\begin{align} \label{eq:hmfenergycurrenttss} 
\left. \d_t \energy^{hmf} (t) \right|_{tss} &= \left. \d_t \energy(t) \right|_{tss} \\
\left. \d_t \entropy^{hmf} (t) \right|_{tss} &= \left. \d_t \entropy (t) \right|_{tss}  \label{eq:hmfentropycurrenttss} . 
\end{align}

Moreover, we conclude that an agreement of the definitions for the time-integrated  quantities would be achieved in the limit of TSS, if the HMF was defined as $\hamiltonian^{hmf^*} ~\equiv~ \freeenergy_{\bm{2}|1}^{eq} $ which corresponds to the definition $\freeenergy_{hmf}^{eq^*} ~\equiv~ \freeenergy^{eq} $. In this case, the equivalence of definitions would still be true for a time-dependent Hamiltonian $\energydensity_2$.

By construction, the definitions of work agree [cf. Eqs. \eqref{eq:twowork} and \eqref{eq:hmfwork}], thus it follows from Eq. \eqref{eq:hmfenergycurrenttss} that the definitions of heat \emph{current} also coincide
\begin{align} \label{eq:twoentropyproductionequivalancetsspartial}
\left. \dot{\heat}^{hmf}(t) \right|_{tss} = \left. \d_t \energy^{hmf}(t) \right|_{tss} - \left. \dot{\work}(t) \right|_{tss} = \left. \dot{\heat}(t) \right|_{tss} . 
\end{align}
Since according to Eqs. \eqref{eq:hmfenergycurrenttss} and \eqref{eq:twoentropyproductionequivalancetsspartial} the entropy production \emph{rates} are also identical,
\begin{align} \label{eq:twoentropyproductionequivalancetsspartialtwo}
\left. \dot{\ep}^{hmf}(t) \right|_{tss} = \left. \d_t \entropy^{hmf}(t) \right|_{tss} - \left. \invtemperature \dot{\heat}^{hmf}(t) \right|_{tss} = \left. \dot{\ep}(t) \right|_{tss} , 
\end{align}
we find that at the differential level the Hamiltonian of mean-force formalism captures the full thermodynamics in the limit of TSS. 
Furthermore, we have proven that in the limit of TSS all definitions of the entropy production rate in Eqs. \eqref{eq:twoentropyproduction}, \eqref{eq:twocoarsegrainedentropyproductionexclusive} and \eqref{eq:hmfsecondlaw} are equivalent, \idest 
\begin{align}
\left. \dot{\ep}(t) \right|_{tss} = \left. \dot{\ep}^{(1)}(t) \right|_{tss} = \left. \dot{\sigma}^{(1)}(t) \right|_{tss} = \left. \dot{\ep}^{hmf}(t) \right|_{tss} .
\end{align}
Together with Eq. \eqref{eq:secondlawfastparticle}, this proves the equality signs in Eq. \eqref{eq:entropyproductionraterelationfull} in the limit of TSS.

This constitutes our first main result:
In the limit of TSS and local equilibrium, the effective thermodynamic descriptions resulting from marginalization and the HMF formalism fully capture the full thermodynamics. In contrast, the effective bipartite description does not match with the full thermodynamics since it neglects the correlations between the two particles.

\subsubsection{Large-Mass Limit: The Work Source}
\label{sec:twoheavyparticle}

We proceed by studying the limit of a diverging mass of the second particle, $\mass_2 \to \infty$, that has already been discussed in Sec. \ref{sec:singlespecialcases}. In view of active Brownian motion, this limit is interesting since the heavy second particle could represent a passive cargo, while the light particle may be considered active. Again, in order to avoid any triviality we assume that the potentials scale with the mass $\mass_2$ as follows: $ \mathit{O}(\D_{\bm{\pos}_{2_i}} \potential_2/ \mass_2) = 1 \, \forall i$ while $ \D_{\bm{\pos}_{2_i}} \potential^{int} / \mass_2 \to 0  \, \forall i$ as $\mass_2 \to \infty$.
Because of the infinite mass of particle two its motion occurs deterministically such that we can neglect the influence of particle one. Consequently, the marginal probabilities become statistically independent and the conditional distribution reads
\begin{align}  \label{eq:conditionalprobabilityheavyparticle}
\probability_{2|1}^{det}(\bm{\coordinate_2},t) = \probability_2^{det}(\bm{\coordinate}_2,t) = \delta(\bm{\pos}_2 - \bm{\pos}_t) \, \delta(\bm{\velocity}_2 - \bm{\velocity}_t) ,
\end{align}
for all times $t$ including the initial time $t=0$.
Here, $\bm{\pos}_t$ and $\bm{\velocity}_t$ are the solutions of the deterministic equations of motion \eqref{eq:singleequationofmotionheavyparticle}. As a result, the effective force \eqref{eq:twonoconservativeforce} becomes conservative, 
\begin{align}  \label{eq:nonconservativeforceheavyparticle} 
\left. \bm{\force}^{(1)}(\bm{\pos}_1,t) \right|_{det} = - \D_{\bm{\pos}_1}  \left.  \potential^{int}_{12}(\bm{\pos}_1,\bm{\pos}_2,t) \right|_{\bm{\pos}_2 = \bm{\pos}_t} ,
\end{align} 
where the notation $\left. Z \right|_{tss}$ corresponds to the conditional probability $\probability_{2|1}$ in the expression $Z$ being substituted by the delta-correlated one in Eq. \eqref{eq:conditionalprobabilityheavyparticle}. Thus, we are dealing with a closed effective Fokker-Planck Eq. \eqref{eq:twocoarsegrainedfokkerplanckequation} for the light particle one that is externally driven by the deterministic motion of the heavy second particle. \\

\paragraph*{Marginalization.}
Since the marginal probabilities are statistically independent, the conditional Shannon entropy \eqref{eq:conditionalshannonentropy} vanishes,
\begin{align} \label{eq:conditionalshannonentropyheavy}
\left. \entropy_{\bm{2}|\bm{1}} \right|_{det} = 0
\end{align}
such that the definition of the effective heat \eqref{eq:twocoarsegrainedheatexclusive} reduces to the naive one \eqref{eq:twocoarsegrainedentropybalanceexclusivenaive}, $
\left. \dot{\heat}^{(1)} \right|_{det}  = \left. \dot{q}^{(1)} \right|_{det} $.
Therefore, by inserting Eq. \eqref{eq:conditionalprobabilityheavyparticle} into Eq. \eqref{eq:twoheatdefintion}, we get
\begin{align} \label{eq:heatheavyparticle}
\left. \dot{\heat}  \right|_{det} - \left. \dot{q}^{(1)} \right|_{det} = \left. \dot{q}^{(2)} \right|_{det} = - \friction_2 \, \bm{\velocity}_t^2 .
\end{align}
Thus, the first law of thermodynamics remains - up to a macroscopic frictional term related to the heavy particle - formally the same as in Eq. \eqref{eq:twofirstlaw},
\begin{align}
\left. \d_t \energy  \right|_{det} = \left. \dot{q}^{(1)} \right|_{det} + \left. \dot{\work}  \right|_{det} - \friction_2 \, \bm{\velocity}_t^2 = \left. \dot{\heat}  \right|_{det} + \left. \dot{\work}  \right|_{det} .
\end{align}
Here, the difference is that the time-dependence of all quantities comes from the dynamical time-dependence of particle one alone, the parametric time-dependence of the Hamiltonian and the deterministic trajectory of the second particle $(\bm{\pos}_t,\bm{\velocity}_t)$.
Further, Eq. \eqref{eq:conditionalshannonentropyheavy} implies that the definitions for the single-particle Shannon entropy \eqref{eq:singleparticleshannonentropy} and the full system entropy agree \eqref{eq:twoaverageentropy},
\begin{align}
\left. \d_t \entropy_{\bm{1}} \right|_{det} = \left. \d_t \entropy \right|_{det} , 
\end{align}
which, in turn, proves that the effective second law of thermodynamics \eqref{eq:twocoarsegrainedentropyproductionexclusive} - up to a macroscopic frictional term of the heavy particle - formally also remains the same as in Eq. \eqref{eq:twoentropyproduction},
\begin{align}  \label{eq:secondlawheavyparticle} 
\left. \dot{\ep}^{(1)} \right|_{det} = \left. \d_t \entropy_{\bm{1}} \right|_{det}  - \left. \invtemperature_1 \dot{q}^{(1)} \right|_{det} = \left. \dot{\tilde{\ep}}  \right|_{det} ,
\end{align}
where $ \left. \dot{\tilde{\ep}}  \right|_{det} = \left. \dot{\ep}  \right|_{det} - \invtemperature_2 \, \friction_2 \, \bm{\velocity}_t^2 $.
The effective thermodynamic description for the two particles therefore reduces, up to a simple macroscopic term, to the standard one of a single particle that is subjected to an external driving. Consequently, the physical interpretation of this limit is that the second particle represents a work source that modulates the energy landscape of the first particle according to a protocol $(\bm{\pos}_t,\bm{\velocity}_t)$. If the deterministic particle is furthermore Hamiltonian, $\friction_2 =0$, the work source is non-dissipative and the effective description coincides with the full one. \\

\paragraph*{Bipartite System.}
Owing to the statistical independence of the marginal distributions, the mutual information \eqref{eq:mutualinformation} and thus the information flow is identically zero, 
\begin{align}  \label{eq:informationrateheavyparticle}
\left. \mutual \right|_{det} = \left. \dot{\mutual}^{2\to1} \right|_{det} = \left. \dot{\mutual}^{1 \to 2} \right|_{det} = 0 .
\end{align}
As a result, the effective entropy balance of the light particle coincides with the full one,
\begin{align}
\left. \d_t \entropy_{\bm{1}} \right|_{det} \! = \! \left. \invtemperature_1 \dot{q}^{(1)} \right|_{det} \!\! + \! \left. \dot{\sigma}^{(1)} \right|_{det} \!=\! \left. \dot{\heat}  \right|_{det} \!\!+\! \left. \dot{\ep}  \right|_{det} \!\!= \left. \d_t \entropy  \right|_{det} ,
\end{align}
while the corresponding effective entropy balance equation for the heavy particle takes the simple macroscopic form
\begin{align}
\left. \invtemperature_2 \, \dot{q}^{(2)} \right|_{det} = \left. - \dot{\sigma}^{(2)} \right|_{det}  = - \invtemperature_2 \, \friction_2 \, \bm{\velocity}_t^2 .
\end{align}  \\

\paragraph*{Hamiltonian of Mean Force.}
The large-mass limit represents a special case of systems away from TSS. Yet, the assumption of a conditional Gibbs state \eqref{eq:hmfconditionalprobability} is inconsistent with the independent single-particle distributions \eqref{eq:conditionalprobabilityheavyparticle}. Therefore, the HMF formalism and the deterministic limit are incompatible.

We can therefore summarize our second main result:
In the deterministic limit, the effective thermodynamics of the first two coarse-graining schemes - marginalization and bipartite structure - are, up to a simple macroscopic frictional term, equivalent to the full thermodynamics. In contrast, the HMF formalism is incompatible with the deterministic limit. In fact, the HMF thermodynamics only matches with the full one in the limit of TSS. This is not surprising since the HMF definitions [cf. Eqs. \eqref{eq:hmfdefinition} and \eqref{eq:hmfconditionalprobability}] are motivated by equilibrium thermostatics.
Notably, in the TSS limit there is a completely asymmetric information flow from the slow to the fast particle, while in the deterministic limit all information flows vanish.

\section{Two Linearly Coupled Harmonic Oscillators}
\label{sec:example}
\subsection{Full Solution}

In this section, the results derived above are illustrated for an analytically solvable example.
For this purpose, we consider an isothermal version of the setup in Fig. \ref{fig:modelschematics} in one dimension. Moreover, the Hamiltonian \eqref{eq:twoenergydensity} is assumed time-independent
\begin{align}  \label{eq:examplepotential}
\potential(\pos_1,\pos_2) = (k_1 \pos_1^2)/2 + (k_2 \pos_2^2)/2 + \invtemperature (\pos_1 \pos_2) ,
\end{align}
and the nonconservative forces $\bm{\force}_i$ are taken as zero. Consequently, there is no work done on or by the two-particle system, $\d_t \energy = \d_t \heat$.
The Fokker-Planck Eq. \eqref{eq:twofokkerplanckequation} reads
\begin{align} \label{eq:examplefokkerplanck}
\D_t \, \probability = - \nabla \cdot ( \bm{\gamma} \cdot \bm{\coordinate}  \probability ) + \nabla^{\top} \cdot \big( \bm{D} \cdot \nabla \probability \big) ,
\end{align}
with $\bm{\coordinate} = (\pos_1,\velocity_1,\pos_2,\velocity_2)^{\top}$ and $ \nabla ~\equiv ~ (\D_{\pos_1},\D_{\velocity_1},\D_{\pos_2},\D_{\velocity_2})^{\top} $.
The constant drift coefficient and diffusion matrix read, respectively,
\begin{align}
\bm{\gamma} &= \begin{pmatrix}
   0 & 1 & 0 & 0 \\ -\frac{k_1}{\mass_1} & -\frac{\friction_1}{\mass_1} & 
   -\frac{\invtemperature}{\mass_1} & 0 \\ 0 & 0 & 0 & 1 \\ -\frac{\invtemperature}{\mass_2 } & 0 & -\frac{k_2}{\mass_2} & -\frac{\friction_2}{\mass_2} 
  \end{pmatrix}
\end{align}
\begin{align}
\bm{D} &= \begin{pmatrix}
  0 & 0 & 0 & 0 \\ 0 & \frac{ \friction_1 }{\invtemperature \mass_1^2} & 0 & 0 \\ 0 & 0 & 0 & 0 \\ 0 & 0 & 0 & \frac{ \friction_2 }{\invtemperature \mass_2^2}
\end{pmatrix}  .
\end{align}
This partial differential equation is supplemented by the initial condition $\probability(0) = \delta(\bm{\coordinate}(t) - \bm{\coordinate}(0) )$.
The solution of this Fokker-Planck equation is given by a Gaussian \cite{risken}
\begin{align} \label{eq:examplefokkerplancksolution}
\probability = \frac{1}{ (2\pi)^2 \, \sqrt{\det \bm{\covariance}} } 
\exp \Big[ \!\! - \! \frac{1}{2} ( \bm{\coordinate} \!-\! \langle \bm{\coordinate} \rangle )^{\top} \! \cdot \! \bm{\covariance}^{-1} \! \cdot \! ( \bm{\coordinate} \! - \! \langle \bm{\coordinate} \rangle ) \Big] ,
\end{align}
where the average values of the coordinates are determined as follows 
\begin{align}  \label{eq:examplefokkerplancksolutionnotationone}
\langle \bm{\coordinate} \rangle (t) = \euler^{ \bm{\gamma} t} \cdot \bm{\coordinate}(0) ,
\end{align}
and the covariance matrix is calculated as
\begin{align}  \label{eq:examplefokkerplancksolutionnotationtwo}
\begin{aligned}
\bm{\covariance}_{kl}(t) & \equiv 2 \sum_{i,j} \frac{1 - \euler^{ -( \lambda_i +\lambda_j ) t }}{ \lambda_i + \lambda_j } \, C_{ij} \, u_i^{(k)} u_j^{(l)} 
%\\ &= \left[ \left( \bm{\coordinate} - \langle \bm{\coordinate} \rangle \right) \left( \bm{\coordinate}(t) - \langle \bm{\coordinate} \rangle \right)^{\top} \right]_{kl} 
.
\end{aligned}
\end{align}
Here, we introduced the transformation matrix
\begin{align}
\bm{C} = \bm{V} \cdot \bm{D} \cdot \bm{V}^{\top}, \quad \bm{V} =  \left(\bm{v}^{(1)},\bm{v}^{(2)},\bm{v}^{(3)},\bm{v}^{(4)}\right) ,
\end{align}
where $\lambda_i$ and $\bm{u}^{(i)}$ ($\bm{v}^{(i)}$) denote the $i$th eigenvale and right (left) eigenvector of the drift coefficient matrix $\bm{\gamma}$,  respectively, \idest
\begin{align}
\begin{aligned}
\bm{\gamma} \cdot \bm{u}^{(i)} &= \lambda_i \, \bm{u}^{(i)}  \\
\bm{v}^{(i)} \cdot \bm{\gamma} &= \lambda_i \, \bm{v}^{(i)} ,
\end{aligned}
\end{align}
such that the left and right eigenvectors of $\bm{\gamma}$ constitute an orthonormal dual basis, $\bm{v}^{(i)} \cdot \bm{u}^{(j)} = \delta_{ij}$.
Substituting Eq. \eqref{eq:examplefokkerplancksolution} into Eqs. \eqref{eq:twoheat} and \eqref{eq:twoentropyproduction},  we obtain for the heat current and the entropy production rate
{\small \begin{align}  \label{eq:exampleheatcurrent}
\dot{\heat} &= \sum_{i=1}^{2} \left[ - \friction_i \Big( \bm{\covariance}_{2i,2i} + \langle\bm{\coordinate}_{2i} \rangle^2 \Big) + \frac{\friction_i }{\invtemperature \mass_i} \right] = \sum_{i=1}^{2} \dot{q}^{(i)} \\
\dot{\ep} &= \sum_{i=1}^{2} \left[ \invtemperature \, \friction_i \Big( \bm{\covariance}_{2i,2i} + \! \langle\bm{\coordinate}_{2i} \rangle^2 \Big) \! - \! 2\frac{\friction_i}{\mass_i} \!+\!  \frac{\friction_i }{\invtemperature \mass_i^2} \bm{\covariance}^{-1}_{2i,2i} \right] \!=\! \sum_{i=1}^{2} \dot{\sigma}^{(i)} , 
\label{eq:exampleentropyproductioncurrent}
\end{align} }
and because of Eq. \eqref{eq:twoentropybalance}
\begin{align}
\d_t \entropy(t) = \sum_{i=1}^{2} \left( \frac{\friction_i }{\invtemperature \, \mass_i^2} \bm{\covariance}^{-1}_{2i,2i} - \frac{\friction_i}{\mass_i}\right) .
\end{align}
In the following, the distribution for particle one $\probability_1(t)$ is needed. The latter is readily determined by marginalizing Eq. \eqref{eq:examplefokkerplancksolution} over the coordinates $\bm{\coordinate}_2$ of the second particle,
\begin{align} \label{eq:examplefokkerplancksolutionmarginalized}
\probability_1 = \frac{1}{2\pi \sqrt{\det \tilde{\bm{\covariance}}} } 
\; \exp \left[ - \frac{1}{2} ( \tilde{\bm{\coordinate}} - \langle \tilde{\bm{\coordinate}} \rangle )^{\top} \cdot \tilde{\bm{\covariance}}^{-1} \cdot ( \tilde{\bm{\coordinate}} - \langle \tilde{\bm{\coordinate}} \rangle ) \right] \!,
\end{align}
with $\tilde{\bm{\coordinate}} = (\pos_1,\velocity_1)^{\top} $ and the inverse of the marginalized covariance matrix $\tilde{\bm{\covariance}}$ that is given by

\begin{widetext}
\begin{align}  
\begin{aligned}
\tilde{\bm{\covariance}}^{-1}_{11} &= \frac{1}{ \left( \bm{\covariance}^{-1}_{34} \right)^2 -  \bm{\covariance}^{-1}_{33} \bm{\covariance}^{-1}_{44} } \left[ \left( \bm{\covariance}^{-1}_{14} \right)^2 \bm{\covariance}^{-1}_{33} - 2 \bm{\covariance}^{-1}_{13} \bm{\covariance}^{-1}_{14} \bm{\covariance}^{-1}_{34} + \bm{\covariance}^{-1}_{11} \left( \bm{\covariance}^{-1}_{34} \right)^2 + \left( \bm{\covariance}^{-1}_{13} \right)^2 \bm{\covariance}^{-1}_{44} - \bm{\covariance}^{-1}_{11} \bm{\covariance}^{-1}_{33} \bm{\covariance}^{-1}_{44} \right] \\\
\tilde{\bm{\covariance}}^{-1}_{12} &= \frac{1}{ \left( \bm{\covariance}^{-1}_{34} \right)^2 \!\!-\!\! \bm{\covariance}^{-1}_{33} \bm{\covariance}^{-1}_{44} } \!\! \left[ \! \bm{\covariance}^{-1}_{14} \bm{\covariance}^{-1}_{24} \bm{\covariance}^{-1}_{33} \!\!-\!\! \bm{\covariance}^{-1}_{14} \bm{\covariance}^{-1}_{23} \bm{\covariance}^{-1}_{34} \!\!-\!\! \bm{\covariance}^{-1}_{13} \bm{\covariance}^{-1}_{24} \bm{\covariance}^{-1}_{34} \!\!+\!\! \bm{\covariance}^{-1}_{12} \left( \bm{\covariance}^{-1}_{34} \right)^2 \!\!+\!\! \bm{\covariance}^{-1}_{13} \bm{\covariance}^{-1}_{23} \bm{\covariance}^{-1}_{44} 
\!\!-\!\! \bm{\covariance}^{-1}_{12} \bm{\covariance}^{-1}_{33} \bm{\covariance}^{-1}_{44} \right]  \\ 
\tilde{\bm{\covariance}}^{-1}_{22} &= \frac{1}{ \left( \bm{\covariance}^{-1}_{34} \right)^2 - \bm{\covariance}^{-1}_{33} \bm{\covariance}^{-1}_{44} } \left[ \left( \bm{\covariance}^{-1}_{24} \right)^2 \bm{\covariance}^{-1}_{33} - 2 \bm{\covariance}^{-1}_{23} \bm{\covariance}^{-1}_{24} \bm{\covariance}^{-1}_{34} + \bm{\covariance}^{-1}_{22} \left( \bm{\covariance}^{-1}_{34} \right)^2 + \left( \bm{\covariance}^{-1}_{23} \right)^2 \bm{\covariance}^{-1}_{44} - \bm{\covariance}^{-1}_{22} \bm{\covariance}^{-1}_{33} \bm{\covariance}^{-1}_{44} \right] .
\end{aligned}
\end{align}
\end{widetext}
Inserting Eq. \eqref{eq:examplefokkerplancksolutionmarginalized} into Eqs. \eqref{eq:twonoconservativeforce} and \eqref{eq:mutualinformationforce}, gives the force contribution to the information flow from particle two to one
\begin{align}  \label{eq:examplemutualinformationforce}
\dot{\mutual}^{(2 \to 1)}_F &= - \frac{\invtemperature}{\mass_1} \Big( \tilde{\bm{\covariance}}_{12}^{-1} \, \bm{\covariance}_{13} + \tilde{\bm{\covariance}}_{22}^{-1} \, \bm{\covariance}_{23} \Big) ,
\end{align}
which can be seen by noting that 
\begin{align}
- \frac{\invtemperature}{\mass_1}  \int \d \bm{\coordinate} \, \probability_1 \, \pos_2 \, \D_{\velocity_1} \probability_{2|1} = \frac{\invtemperature}{\mass_1}  \int \d \bm{\coordinate} \, \probability \, \pos_2 \, \D_{\velocity_1} \ln \probability_1   .
\end{align}
Moreover, from Eq. \eqref{eq:twocoarsegrainedentropyproductionexclusive} it follows for the effective entropy production rate that
\begin{align}
\dot{\ep}^{(1)} \!&=\! \invtemperature \, \friction_1 \Big( \! \tilde{\bm{\covariance}}_{22} \!+\! \langle\bm{\coordinate}_{2} \rangle^2 \! \Big) \!-\! 2\frac{\friction_1}{\mass_1} \!+\!  \frac{\friction_1 }{\invtemperature \mass_1^2} \tilde{\bm{\covariance}}^{-1}_{22}  ,
\label{eq:exampleentropyproductioncurrenteffective}
\end{align} 
from which via Eqs. \eqref{eq:twoentropyproductionrewritten} and \eqref{eq:exampleentropyproductioncurrent} we get the entropic contribution to the information flow
\begin{align}  \label{eq:examplemutualinformationentropy}
\dot{\mutual}^{(2 \to 1)}_S \! &= \! \invtemperature \, \friction_1 \Big(  \tilde{\bm{\covariance}}_{22} \!-\! \bm{\covariance}_{22} \Big) \!+\!  \frac{\friction_1 }{\invtemperature \mass_1^2} \Big( \tilde{\bm{\covariance}}^{-1}_{22} \!-\! \bm{\covariance}^{-1}_{22} \Big)  . 
\end{align}
Combining the last three equations with Eqs. \eqref{eq:twocoarsegrainedentropybalanceexclusive}, \eqref{eq:twocoarsegrainedheatexclusive} and \eqref{eq:conditionalshannonentropy}, yields

\begin{widetext}
\begin{align}
\dot{\mutual}^{(2 \to 1)} &= \invtemperature \, \friction_1 \Big(  \tilde{\bm{\covariance}}_{22} - \bm{\covariance}_{22} \Big) +  \frac{\friction_1 }{\invtemperature \mass_1^2} \Big( \tilde{\bm{\covariance}}^{-1}_{22} - \bm{\covariance}^{-1}_{22} \Big) - \frac{\invtemperature}{\mass_1} \Big( \tilde{\bm{\covariance}}_{12}^{-1} \, \bm{\covariance}_{13} + \tilde{\bm{\covariance}}_{22}^{-1} \, \bm{\covariance}_{23} \Big) \\
\d_t \, \entropy_{\bm{2}|\bm{1}}
% &= \invtemperature \, 	\dot{q}^{(2)} + \dot{\sigma}^{(2)} - \dot{\mutual}^{(2 \to 1)} \\
&= \frac{\friction_2}{\invtemperature \mass_2^2} \bm{\covariance}^{-1}_{44} - \frac{\friction_2}{\mass_2}  - \frac{\invtemperature}{\mass_1} \Big( \tilde{\bm{\covariance}}_{12}^{-1} \, \bm{\covariance}_{13} + \tilde{\bm{\covariance}}_{22}^{-1} \, \bm{\covariance}_{23} \Big) - \invtemperature \, \friction_1 \Big(  \tilde{\bm{\covariance}}_{22} - \bm{\covariance}_{22} \Big) - \frac{\friction_1 }{\invtemperature \mass_1^2} \Big( \tilde{\bm{\covariance}}^{-1}_{22} - \bm{\covariance}^{-1}_{22} \Big) \\
\dot{\heat}^{(1)} &= \frac{\friction_2}{\invtemperature \mass_2^2} \bm{\covariance}^{-1}_{44} + \frac{\friction_1}{\mass_1} - \frac{\friction_2}{\mass_2}  - \frac{\invtemperature}{\mass_1} \Big( \tilde{\bm{\covariance}}_{12}^{-1} \, \bm{\covariance}_{13} + \tilde{\bm{\covariance}}_{22}^{-1} \, \bm{\covariance}_{23} \Big) - \invtemperature \, \friction_1 \Big(  \tilde{\bm{\covariance}}_{22} + \langle \bm{\coordinate}_{2} \rangle^2 \Big) - \frac{\friction_1 }{\invtemperature \mass_1^2} \Big( \tilde{\bm{\covariance}}^{-1}_{22} - \bm{\covariance}^{-1}_{22} \Big)  .
\end{align}
\end{widetext}
\phantom{ }

\subsection{Fast-Dynamics Limit}

Since $\bm{\force}_2 = 0$, the limit of TSS implies that the second particle is at local equilibrium conditioned on the coordinates of particle one. Within TSS, the effective force \eqref{eq:equilibriumconservativeforce} reads
\begin{align}  \label{eq:exampleequilibriumconservativeforce}
\force^{(1)}  = \invtemperature^2 \frac{\pos_1}{k_2} , 
\end{align}
and closes the effective Fokker-Planck Eq. \eqref{eq:twocoarsegrainedfokkerplanckequation},
\begin{align} 
\D_t \probability_1 
&= - \nabla_1 \cdot \left[ \left( \bm{\gamma}_1 \cdot \tilde{\bm{\coordinate}} \right) \probability_1 \right] + \nabla_1 \cdot \big( \bm{D}_1 \cdot \nabla_1 \probability_1 \big)  ,
\label{eq:examplefastfokkerplanckequationvectorial}
\end{align}
with $ \nabla_1 ~\equiv ~ (\D_{\pos_1},\D_{\velocity_1})^{\top} $.
The drift coefficient and the diffusion matrix read
\begin{align}
\bm{\gamma}_1 \!=\! \begin{pmatrix}
0 & 1 \\ -\frac{k_1 }{\mass_1} - \frac{ \invtemperature^2 }{k_2  \, \mass_1}  & -\frac{\friction_1}{\mass_1}  \\ \end{pmatrix} \!\! , \quad
\bm{D}_1 \!=\! \begin{pmatrix}
0 & 0 \\ 0 & \frac{ \friction_1 }{\invtemperature \mass_1^2} , \; 
\end{pmatrix}  .
\end{align}
This Fokker-Planck equation implies that we are dealing with a bivariate Ornstein-Uhlenbeck process, thus its solution is given by a bivariate Gaussian \cite{risken}
\begin{align} \label{eq:examplefokkerplancksolutionfast}
\probability_1 = \frac{1}{ 2\pi \, \sqrt{\det \tilde{\bm{\covariance}}} } \; \euler^{ -\frac{1}{2} \left( \tilde{\bm{\coordinate}} - \langle \tilde{\bm{\coordinate}} \rangle \right)^{\top} \cdot \, \tilde{\bm{\covariance}}^{-1} \cdot \, \left( \tilde{\bm{\coordinate}} - \langle \tilde{\bm{\coordinate}} \rangle \right) } \, ,
\end{align}
where the covariance matrix $\tilde{\bm{\covariance}}$ is specified by Eq. \eqref{eq:examplefokkerplancksolutionnotationtwo} and the averages of the coordinates $\tilde{\bm{\coordinate}}$ are determined as follows
\begin{align} 
\langle \tilde{\bm{\coordinate}} \rangle (t) = \euler^{\bm{\gamma}_1 t} \cdot \tilde{\bm{\coordinate}}(0) .
\end{align}

In the following of this subsection, we employ the numerical values $\friction_1= 0.8$, $\invtemperature=0.05$, $k_1 = 1, \mass_1=1$, while we consider three different spring constants $k_2$ masses $\mass_2$ and friction coefficients $\friction_2$: $(k_2 = 15, \mass_{2,a}=5 \,,\, \friction_{2,a}=0.75)$, $(k_2 = 25,\mass_{2,b}=7.5 \,,\, \friction_{2,b}=0.25)$ and $(k_2 = 50,\mass_{2,c}=10 \,,\, \friction_{2,c}=0.1)$.
This choice of parameters corresponds to an increasing separation of the time-scales between the different stochastic dynamics of the two particles. In the order $a-b-c$, the second particle approaches equilibrium conditioned on the coordinates of the first particle: Since the interaction potential scales linearly in the inverse temperature [Eq. \eqref{eq:examplepotential}], we chose a relatively small value for $\invtemperature$ to implement a weak-coupling condition between the first and second particle - a crucial requisite for the second particle to behave like an ideal heat reservoir \cite{breuer2006,lindenberg1990}. As $k_2$ and $\mass_2$ increases and $\friction_2$ decreases, the relaxation time-scale of the second particle further shrinks, hence the time-scales of the particles dynamics start to separate, as desired.
Moreover, we prepare the initial condition \eqref{eq:hmfconditionalprobability} with
$\probability_1(0) = \delta(\tilde{\bm{\coordinate}} - \tilde{\bm{\coordinate}}(0) )$ with $ \tilde{\bm{\coordinate}}(0) = (2,1)^{\top} $.

\begin{figure}[h!] 
\begin{center}

\includegraphics[scale=1]{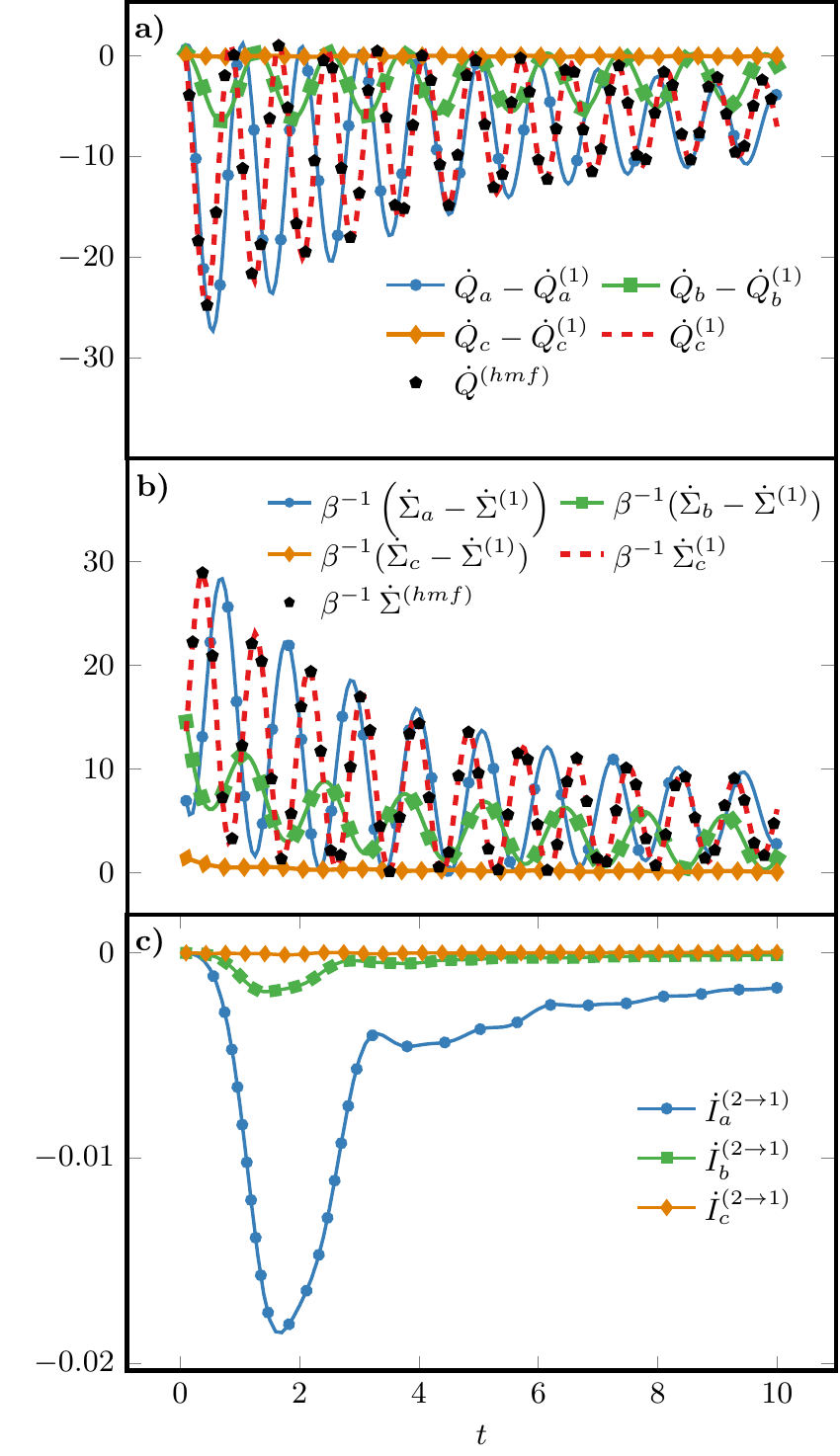}

\caption{Difference between the full $\dot{\heat}$ and effective heat current $\dot{\heat}^{(1)}$ in a) and between the scaled full $\invtemperature \, \dot{\ep}$ and scaled effective entropy production rate $\invtemperature \, \dot{\ep}^{(1)}$ in b) as a function of time $t$. In these two figures, the parameter sets $a$, $b$ and $c$ correspond to the solid blue line with circle markers, the solid green line with square markers and the solid orange line with diamond markers, respectively. Moreover, the effective quantities for the parameters $c$ and those based on the HMF are overlaid in panels a) and b) and correspond to the red dashed line and the black pentagon markers, respectively.
The information flow $\dot{\mutual}^{(2 \to 1)}$ as a function of time and for the parameter sets $a$ (solid blue line with circle markers), $b$ (solid green line with square markers) and $c$ (solid orange line with diamond markers) is depicted in panel c).} \label{fig:examplefastheatcurrententropyproduction}
\end{center}
\end{figure}

Fig. \ref{fig:examplefastheatcurrententropyproduction} depicts in a) the difference between the global $\dot{\heat}$ and effective heat current $\dot{\heat}^{(1)}$ and in b) the scaled difference between the global $\dot{\ep}$ and effective entropy production rate $\dot{\ep}^{(1)}$ as a function of time $t$.
We observe that both the effective heat current and entropy production rate converge to the corresponding full quantities in the limit of TSS. The overall system remains out-of-equilibrium as reflected by finite (effective) heat currents and (effective) entropy production rates of the first particle. Since the corresponding single-particle definition for the heat, $\dot{q}^{(1)}$, does not agree with definition of the effective one [not shown in a)], it follows that the time-derivative of the conditional Shannon entropy, $\d_t \entropy_{\bm{2}|\bm{1}}$ remains finite in the limit of TSS.
We furthermore note that the effective heat current and entropy production rate
are in agreement with the time-derivative of the heat \eqref{eq:hmfheatrewritten} and entropy production \eqref{eq:hmfsecondlaw} using the HMF formalism.
Moreover, Fig. \ref{fig:examplefastheatcurrententropyproduction} c) shows that the directional information flow $\dot{\mutual}^{(2 \to 1)}$ vanishes in the limit of TSS. This in turn implies first that the additive contribution $\dot{\sigma}^{(2)}$ to the full entropy production rate becomes zero while the inverse information flow $\dot{\mutual}^{(1 \to 2)}$ remains finite. It furthermore follows from the nonpositivity of $\dot{\mutual}^{(2 \to 1)}$ that the non-positive entropic contribution $\dot{\mutual}^{(2 \to 1)}_S$ dominates over the non-negative force contribution $\dot{\mutual}^{(2 \to 1)}_F$.

\subsection{Large-Mass Limit}

In the large-$\mass_2$ limit, the effective force \eqref{eq:nonconservativeforceheavyparticle} reads
\begin{align}  \label{eq:exampleheavynonconservativeforce}
\force^{(1)} = \left. - \invtemperature \, \pos_2 \right|_{\pos_2 = \pos_t} ,
\end{align}
and closes the effective Fokker-Planck Eq. \eqref{eq:twocoarsegrainedfokkerplanckequation},
\begin{align} 
\D_t \probability_1 
&= - \nabla_1 \cdot \left[ \left( \bm{\gamma}_1 \cdot \tilde{\bm{\coordinate}} + \bm{\force}^{(1)} \right) \probability_1 \right] + \nabla_1 \cdot \big( \bm{D}_1 \cdot \nabla_1 \probability_1 \big) .  \label{eq:exampleheavyfokkerplanckequationvectorial}
\end{align}
The constant drift coefficient, the scaled effective force vector and the diffusion matrix read
\begin{align}
\bm{\gamma}_1 = \begin{pmatrix}
   0 & 1 \\ -\frac{k_1}{\mass_1} & -\frac{\friction_1}{\mass_1}  \\ 
  \end{pmatrix} , \; 
\bm{\force}^{(1)} = \begin{pmatrix}
  0 \\ - \frac{\invtemperature x_t}{\mass_1 }
\end{pmatrix} , \; 
\bm{D}_1 = \begin{pmatrix}
  0 & 0 \\ 0 & \frac{ \friction_1 }{\invtemperature \mass_1^2} , \;
\end{pmatrix}  .
\end{align}
This partial differential Eq. is supplemented by the initial condition $\probability_1(0) = \delta(\tilde{\bm{\coordinate}} - \tilde{\bm{\coordinate}}(0))$ with $\tilde{\bm{\coordinate}}(0) = (2,1)^{\top} $.
%It can be shown that the solution to this Ornstein-Uhlenbeck process with an additive modulation $\bm{\force}^{(1)}$ is again given by a bivariate Gaussian \cite{jung93,haenggi1975},
%\begin{align} \label{eq:examplefokkerplancksolutionheavy}
%\probability_1 = \frac{1}{ 2\pi \, \sqrt{\det \tilde{\bm{\covariance}}} } \; \exp\left[ -\frac{1}{2} \left( \tilde{\bm{\coordinate}} - \langle \tilde{\bm{\coordinate}} \rangle \right)^{\top} \cdot \, \tilde{\bm{\covariance}}^{-1} \cdot \, \left( \tilde{\bm{\coordinate}} - \langle \tilde{\bm{\coordinate}} \rangle \right) \right] ,
%\end{align}
The averages are determined as follows
\begin{align}  \label{eq:examplefokkerplancksolutionnotationoneheavy}
\langle \tilde{\bm{\coordinate}} \rangle (t) = \euler^{\bm{\gamma}_1 t} \cdot \tilde{\bm{\coordinate}}(0)  + \int_0^t \euler^{\bm{\gamma}_1 (t-t')} \cdot \bm{\force}^{(1)}(t') \, \d t' ,
\end{align}
while the coordinates ($\pos_t,\velocity_t$) of the second particle follow the solution of the deterministic equation of motion \eqref{eq:singleequationofmotionheavyparticle},
\begin{align}  \label{eq:exampleheavydeterministicsolution}
\begin{aligned}
\pos_t &= 2 \cos \left( \frac{k_2}{\mass_2} t \right) + \frac{\mass_2}{k_2} \sin \left( \frac{k_2}{\mass_2} t \right) \\ 
\velocity_t &= \cos \left( \frac{k_2}{\mass_2} t \right) - 2 \frac{k_2}{\mass_2}  \sin \left( \frac{k_2}{\mass_2} t \right) ,
\end{aligned}
\end{align}
for the initial condition as chosen above.
In the following, we employ the numerical values $\friction_1= 0.3$, $\friction_2 = 1.5$, $\invtemperature=1$, $k_1 = 4\,,\, \mass_1=1$, while we consider three different masses $\mass_2$ and constants $k_2$ such that their ratio remains constant: $(\mass_{2,a}=4 \,,\, k_{2,a}=3.8)$, $(\mass_{2,b}=40 \,,\, k_{2,b}=38)$ and $(\mass_{2,c}=400 \,,\, k_{2,c}=380)$.
It is important to note that the set of parameters $a,b,$ and $c$ is chosen such that the ratio of $\mass_2$ and $k_2$ remains constant and thus leaves the deterministic trajectory of the second particle invariant according to Eq. \eqref{eq:exampleheavydeterministicsolution}.

\begin{figure}[h!] 
\begin{center}

\includegraphics[scale=1]{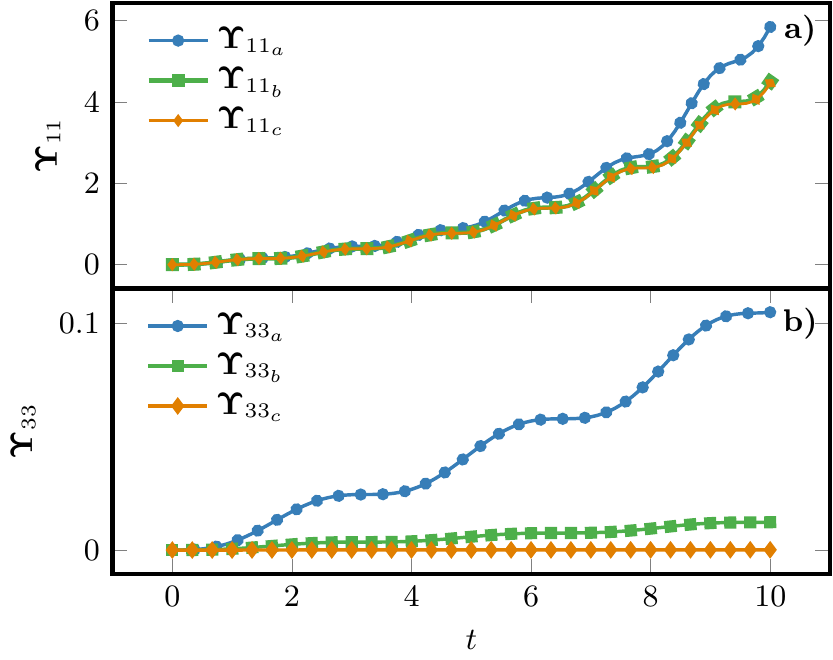}

\caption{Variance $\bm{\covariance}_{11}$ in a) and $\bm{\covariance}_{33}$ in b) of the positional degrees of freedom $\pos_1$ and $\pos_2$, respectively, as a function of time $t$ and for the different parameter sets $a$ (solid blue line with circle markers), $b$ (solid green line with square markers) and $c$ (solid orange line with diamond markers). \label{fig:examplepositionvariance}  }
\end{center}
\end{figure}

Fig. \ref{fig:examplepositionvariance} depicts the variances $\bm{\covariance}_{11}$ and $\bm{\covariance}_{33}$ of the positional variables $\pos_1$ and $\pos_2$, in panels a) and b) respectively. As expected, the fluctuations of the first particle do not exhibit striking qualitative changes since the variance of the second particle vanishes with growing mass $\mass_2$. We verify that
\begin{align}
\bm{\covariance}_{ij} = 0 , \quad \forall \, ij \neq \lbrace 11,12,21,22 \rbrace
\end{align}
thus confirming that the second particle behaves deterministically in the large-$\mass_2$ limit as prescribed by the equations of motion \eqref{eq:exampleheavydeterministicsolution}.

\begin{figure}[h!] 
\begin{center}

\includegraphics[scale=1]{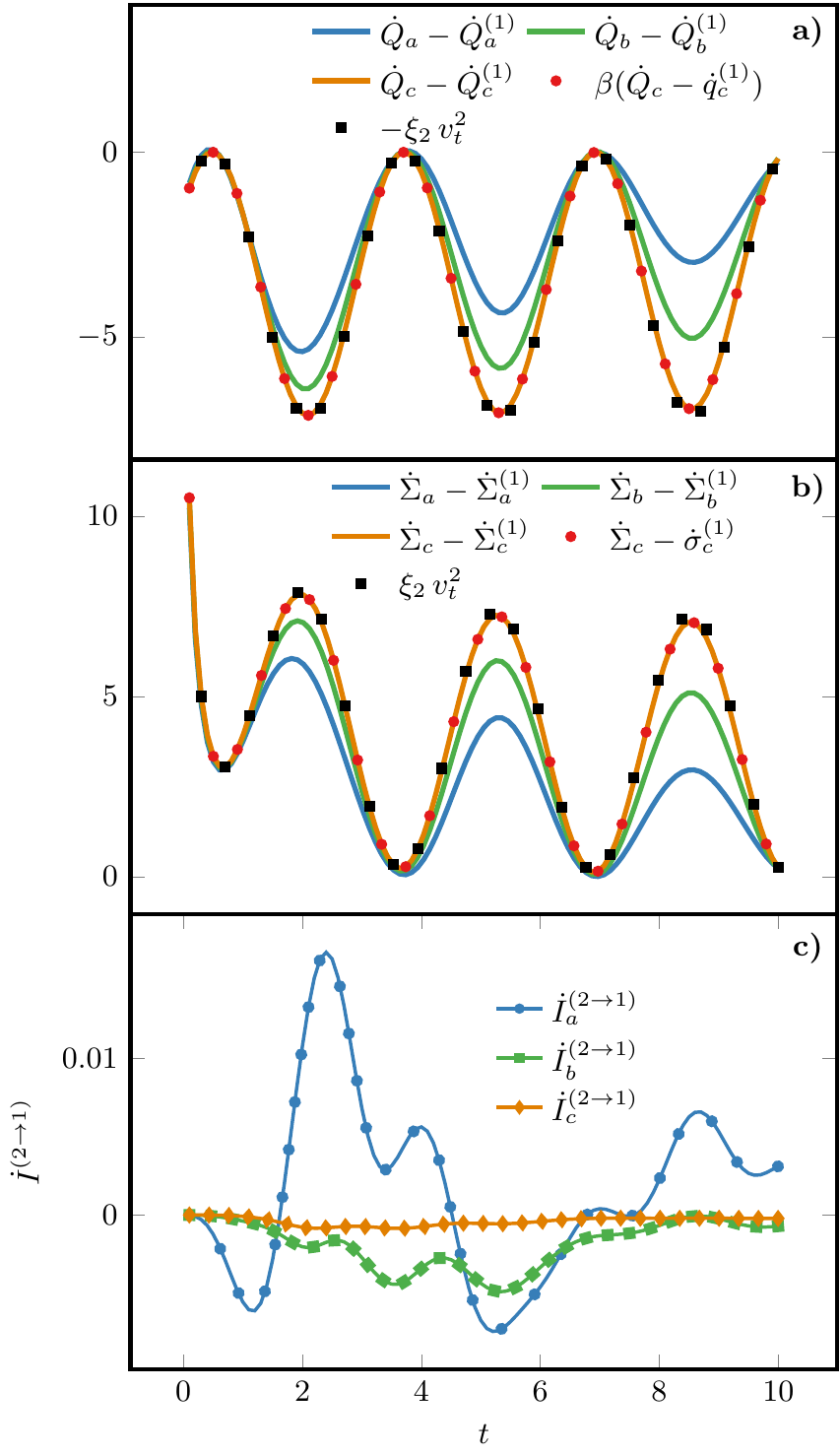}

\caption{Difference between the full $\dot{\heat}$ and effective heat current $\dot{\heat}^{(1)}$ in a) and between the full $\dot{\ep}$ and effective entropy production rate $\dot{\ep}^{(1)}$ in b) as a function of time $t$.
In panel a) [b)], the parameter sets $a$, $b$ and $c$ correspond to the upper [lower] blue, intermediate green and lower [upper] orange line, respectively. Moreover, the difference between the full $\dot{\heat}$ \big[$\dot{\ep}$\big] and the naive definition of the effective heat current [entropy production rate] $\dot{q}^{(1)}$ \big[$\dot{\sigma}^{(1)}$\big] is overlaid in panel a) [b)]. The information flow $\dot{\mutual}^{(2 \to 1)}$ as a function of time and for the parameter sets $a$ [solid blue line with circle markers], $b$ [solid green line with square markers] and $c$ [solid orange line with diamond markers] is depicted in panel c).
\label{fig:exampleheavyheatcurrent} }
\end{center}
\end{figure}

Next, Fig. \ref{fig:exampleheavyheatcurrent} a) shows that the effective heat current, $\dot{\heat}^{(1)}$, converges to the full one, $\dot{\heat}$, minus the macroscopic dissipation of the heavy particle, $ \friction_2 \velocity_t^2$, as $\mass_2$ increases.

This macroscopic term is naturally non-negative and periodic with the frequency $k_2/\mass_2$ due to the choice of a harmonic potential \eqref{eq:examplepotential}.
Furthermore, Fig. \ref{fig:exampleheavyheatcurrent} b) illustrates the convergence of the effective entropy production $\dot{\ep}^{(1)}$ to the full one, $\dot{\ep}$, plus the macroscopic dissipation of the heavy particle with increasing $,\mass_2$. Since the single-particle definitions for the heat current, $\dot{q}^{(1)}$, and the entropy production rate, $\dot{\sigma}^{(1)}$, also converge to the full quantities, respectively, it follows that the time-derivative of the conditional Shannon entropy, $\d_t \entropy_{\bm{2}|\bm{1}}$, and the information flow from the light to the heavy particle, $\dot{\mutual}^{(1\to2)}$, vanish as $\mass_2$ grows.

Finally, in Fig. \ref{fig:exampleheavyheatcurrent} c) the directional information flow from the heavy to the light particle $\dot{\mutual}^{(2\to1)}$ is shown to decrease in modulus with increasing $\mass_2$. It is interesting to note that the vanishing directional flow becomes negative if $\mass_2$ is sufficiently large. This means that the non-positive entropic contribution converges at a slower rate to zero than the force one does.

\section{Conclusion}
\label{sec:conclusion}

In this work, we presented three coarse-graining approaches for the thermodynamics of two interacting underdamped Brownian particles:
The observation of only one particle while the other one has been coarse-grained, the partitioning of the two-body system into two single-particle systems exchanging information flows and the Hamiltonian of mean force formalism.
We demonstrated that the effective thermodynamics of the first and third approach is equivalent to the correct global thermodynamics in the limit of time-scale separation between the two particles, where the faster evolving particle equilibrates with respect to the coordinates of the more slowly evolving particle.
Conversely, we observed a mismatch between the effective and full thermodynamics in the bipartite case, since the entropic contribution due to the coupling of the two particles is not taken into account.
Physically, in this limit the faster evolving particle becomes part of the heat reservoir to which the other particle is coupled.
Conversely, if one particle becomes deterministic because of an exceedingly large mass compared to the other particle's mass, it acts as an additional work source on the lighter particle. In this case, the effective thermodynamics of the first two of the aforementioned three approaches agree, up to a simple macroscopic term related to the dissipation of the work source, with the correct global one.
The Hamiltonian of mean force formalism however was shown to be incompatible with the large-mass limit. In fact, the same is true for any physical regime outside the time-scale separation limit.
This reflects that the Hamiltonian of mean force formalism was originally motivated by and employed in equilibrium thermostatics. These theoretical predictions were confirmed via an analytically tractable model made up of two linearly coupled harmonic oscillators. 
We remark that the generalization to an arbitrary many-body system, where particle one and two are replaced by two subsets of interacting particles is straightforward. Since the findings for systems with arbitrarily many bodies are identical to the results for the two-body setup reported above, an explicit presentation of the former is omitted.

In view of applications, we remark that due to the emergence of a velocity-dependent nonconservative force in the effective description, our work could be of significant methodological value for active particles, where the issue of coarse-graining is indeed very important. To conclude, we leave the study of effective fluctuating thermodynamics in underdamped systems for future works. In this context, it would also be interesting to explore if other coarse-graining schemes that were applied to jump processes, for instance as proposed in Refs. \cite{polettini2017prl,polettini2019jsp}, can also be utilized in underdamped Fokker-Planck systems. These studies could give rise to new strategies for systematically and thermodynamically consistently coarse-graining many-body systems. \\

\section*{Acknowledgments}

We gratefully acknowledge funding by the National Research Fund, Luxembourg (AFR PhD Grant 2016, No. 11271777 and ATTRACT Project No. FNR/A11/02) and by the European Research Council (Project NanoThermo, ERC-2015-CoG Agreement No. 681456).

\bibliography{bibliography}

\end{document}